\documentclass[10pt,twocolumn]{IEEEtran}
\usepackage{amsfonts}
\usepackage{indentfirst}
\usepackage{amsmath}
\usepackage{amsthm}
\usepackage{amssymb}
\usepackage{color}
\usepackage{graphicx}
\usepackage[para]{threeparttable}
\usepackage{booktabs}
\usepackage{makecell}
\usepackage{diagbox}
\usepackage{multirow}
\usepackage{epstopdf}
\usepackage{mathrsfs}
\usepackage{bm}
\usepackage{subfigure}
\usepackage{multicol}
\usepackage{stfloats}
\usepackage{cite}

\theoremstyle{remark}

\newtheorem{Theorem}{\textit{Theorem}}
\newtheorem{Lemma}{\textit{Lemma}}
\newtheorem{Remark}{\textit{Remark}}

\usepackage[ruled,linesnumbered]{algorithm2e}
\allowdisplaybreaks[4]

\usepackage[colorlinks=true,linkcolor=blue,anchorcolor=blue,urlcolor=blue,citecolor=blue]{hyperref}%
\renewcommand{\eqref}[1]{
	\textcolor{blue}{(}
	\hyperref[#1]{\ref*{#1}}
	\textcolor{blue}{)}
}

\begin{document}
	\title{MIMO Multifunction Radar-Communication System: Waveform and Filter Joint Design}
	\author{Yu Zhou, $ \text{Zhengchun Zhou}^{*} $, \IEEEmembership{Senior Member, IEEE}, Qiao Shi, \IEEEmembership{Member, IEEE},\\ Zilong Liu, \IEEEmembership{Senior Member, IEEE}, and Pingzhi Fan, \IEEEmembership{Life Fellow, IEEE}
		\thanks{Yu Zhou, Zhengchun Zhou and Qiao Shi are with the School of Information Science and Technology, Southwest Jiaotong University, Chengdu 611756, China (e-mail: 2021201698@my.swjtu.edu.cn; zzc@swjtu.edu.cn; qiaoshi@swjtu.edu.cn). (\emph{Corresponding author: Zhengchun Zhou}.)}
		\thanks{Zilong Liu is with the School of Computer Science and Electronics Engineering, University of Essex, Colchester CO4 3SQ, U.K. (e-mail: zilong.liu@essex.ac.uk).}
		\thanks{Pingzhi Fan is with the Key Laboratory of Information Coding and Wireless Communications, Southwest
			Jiaotong University, Chengdu 611756, China (e-mail: pzfan@swjtu.edu.cn).}}
	\maketitle
	\begin{abstract}
		This paper studies a novel multiple-input-multiple-output multifunction radar-communication (MIMO MRC) system that simultaneously performs downlink communications, electronic jamming and target detection in the presence of signal-dependent interference sources.	 
		We formulate a joint transmit waveform and receive filter design optimization problem under total energy (TE) and TE plus peak-to-average power ratio (TE+PAPR) constraints, respectively.
		Specifically, a jamming functionality constraint is introduced to jam electronic threat (ET) targets.
		To ensure the capabilities of both detection and communications, we maximize the radar receive signal-to-interference-noise ratio (SINR) while minimizing communication multiuser interference (MUI) energy under a directional target energy constraint. 	
		Moreover, we also consider the practical challenge of the imperfect communication channel state information (CSI). 
		A computationally efficient gradient projection framework-based alternating direction method of multipliers (GP-ADMM) algorithm is proposed to solve the formulated challenging non-convex problem. It is proved that the proposed the algorithm converges to a Karush-Kuhn-Tucker (KKT) stationary point.
		Simulation results demonstrate the effectiveness of the proposed MIMO MRC system in achieving multifunctional operation, showing significant performance in simultaneous target detection, reliable communications, and electronic jamming. 
	\end{abstract}
	
	
	\section{Introduction}\label{section1}
    
	\subsection{Background and Related Work}
	\IEEEPARstart{T}{he} sixth generation (6G) of wireless networks is envisioned as a key enabler for numerous emerging applications, including autonomous driving, smart manufacturing, digital twins, and the low-altitude economy \cite{M. Chafii-2023}. Integrated sensing and communication (ISAC) is considered as one of the key technologies for 6G wireless networks, offering enhanced sensing and communication capabilities \cite{A. Kaushik-2024}. Unlike traditional approaches that treat communications and sensing as separate systems, ISAC technology yields higher spectrum utilization and lower hardware costs by sharing the spectrum and hardware resources between sensing (e.g., radar) and communications systems \cite{F. Liu-2022}, \cite{X. Luo-2025}. Therefore, ISAC is attracting growing research attention from both the academic and industrial communities \cite{F. Liu-2023}, \cite{Z. Wei-2025}.\par
	
	Nonetheless, the ISAC design faces several challenges. From a hardware standpoint, radar and communications systems often have distinct requirements, particularly in the operation region of power amplifiers \cite{C. Sturm-2011}, where radar demands higher peak power. Besides, from the signal processing perspective, efficient design of ISAC waveforms that cater to different application scenarios under limited wireless resources remains a key challenge \cite{S. Lu-2024}. In order to meet the key performance indicators for the different use cases, multiple-input multiple-output (MIMO) technology has been widely employed in ISAC, as it allows us to fully exploit the spatial degrees of freedom to optimize the radar-communications performance trade-off \cite{X. Liu-2020}. Therefore, a number of ISAC waveform design efforts have been dedicated to beamforming \cite{R. P. Sankar-2024}, \cite{Y. Xu-2025} and precoding techniques \cite{S. Lu-2024-Random}.
	\par

    Specifically, in \cite{F. Liu-2018-2}, the ISAC waveform was optimized by minimizing multi-user interference (MUI), addressing both omnidirectional and directional beamforming design. To further enhance radar beamforming performance, the joint design of transmit ISAC waveforms and receive filters was investigated in \cite{C. G. Tsinos-2021}, by maximizing the radar receiver signal-to-interference-noise ratio (SINR).  \cite{C. Jia-2025} investigated joint waveform and receive filter optimization in a multi-user and multi-target MIMO ISAC system.
    It is shown in \cite{R. Liu-2021} that the constructive interference (CI) can push the received communication signal toward the correct decision region, thereby further enhancing communication quality of service (QoS). In \cite{C. Huang-2024}, the radar output SINR was maximized as a cost function, while formulating communication CI constraints with imperfect channel state information (CSI). 
    To fully utilize the spatial degree of freedom (DoF) gain at the transmitter,  a joint design of the ISAC linear precoder and the radar receive beamformer was proposed in \cite{C. G. Tsinos-2024}, thus aiming to optimize the performance of the radar part while maintaining a desired QoS for the communication.
    Further, \cite{Z. -T. Liao-2025} proposed a hybrid beamforming design for MIMO orthogonal frequency division multiplexing (OFDM) ISAC systems by studying
    the impact of beamforming on the sensing range-Doppler (RD) map and system capacity.   
    To evaluate the target estimation capabilities of ISAC systems, a beamformer was designed by minimizing the Cramér-Rao bound (CRB) of the target estimation while guaranteeing the communication SINR \cite{F. Liu-2022-CRB}. A bounded direction-of-arrival (DoA) error model and a statistical DoA error model were considered in \cite{Z. Xing-2025}, and beams were designed by optimizing the worst-case CRB to address degradation of performance due to DoA estimation errors.\par

	Although the aforementioned ISAC waveform designs achieve improved performance through joint optimization of radar and communication metrics, they mostly assume oversimplified radar and communications environments, such as interference-free or clutter interference, and so on.
	In the adversarial environment, the presence of electronic threat (ET) targets (e.g., eavesdroppers or hostile radars) poses a significant threat to ISAC systems \cite{H. Guo-2022}. On the one hand, as potential eavesdroppers, ET targets may intentionally intercept critical information from signals leaked by the ISAC system \cite{X. Li-2021}. On the other hand, hostile radars can significantly enhance the enemy battlefield awareness and electronic warfare capabilities by continuously monitoring and collecting environmental information \cite{T. Gulum-2017}.\par
	
	Recently, physical layer security (PLS) has been recognized as a promising solution to counter malicious eavesdroppers \cite{Z. Wei-2022}, \cite{M. Liu-2025}. Several efforts have been made to achieve secure communications under eavesdropping scenarios.
	In \cite{N. Su-2021} and \cite{F. Dong-2023}, the SINR of eavesdroppers was suppressed through beamforming and artificial noise (AN) while guaranteeing the performance of radar and legitimate communications.
	To address the scenario where the CSI of eavesdroppers is imperfectly known, a robust design of communication and radar beamforming was proposed that accounts for both bounded and Gaussian CSI errors of eavesdroppers \cite{Z. Ren-2023}. These efforts have focused on introducing AN at the transmitter and designing a transmit precoding matrix. In \cite{D. Li-2024}, the transmit energy at each eavesdropper direction was limited to a low level such that the probability of being intercepted was significantly reduced. Besides, a novel sensing-aided secure communication scheme was proposed to jointly control the performance of radar and communication \cite{Y. Cao-2025}.
    However, the aforementioned works primarily focus on mitigating eavesdropping threats, leaving other potential threats unaddressed, e.g., threats related to battlefield awareness. 
	Notably, it is of interest to design a multifunction radio frequency (MFRF) system that integrates radar, communications, and electronic jamming functions via a shared aperture architecture, radio frequency (RF) components, and unified signal processing, thus dramatically improving the electronic countermeasures (ECM) capability \cite{G. C. Tavik-2005}. 
	In \cite{B. Tang-2022}, a MIMO MFRF system was presented, and its achievable SINR was analyzed.
	The findings highlight the significant capability of MIMO arrays to support multiple functions simultaneously. Furthermore, \cite{D. Li-2024-RIS} investigated the waveform design problem for reconfigurable intelligent surface (RIS)-assisted MFRF systems, thereby optimizing relative entropy to enhance target detection performance.\par
	
	\subsection{Motivations and Contributions}
   
    The MIMO MFRF systems in \cite{B. Tang-2022} and \cite{D. Li-2024-RIS} exploit the directional gain of MIMO arrays to transmit jamming signals toward ET targets and communication signals toward users, respectively. However, when an ET target is located close to a communication user in the angular domain, the spatial correlation between their channel responses makes it challenging to simultaneously satisfy the jamming and communication requirements through conventional direction-based spatial separation. Furthermore, for multi-target detection in MIMO ISAC systems, existing studies have predominantly focused on angular-domain beamforming, while the effects of target and interference responses in different range bins have not been sufficiently investigated in some related designs. Ignoring the range-domain characteristics may result in degraded target detection performance, particularly in the presence of signal-dependent interference.
    Motivated by these challenges, this paper proposes a novel MIMO multifunction radar-communication (MIMO MRC) system that explicitly accounts for targets and signal-dependent interference located in different range bins. The proposed system aims to jointly achieve multi-target detection in the presence of interference sources, communication with multiple users, and jamming against ET targets. The main contributions of this work are summarized as follows.

	\begin{itemize}
	\item We propose a joint transmit waveform and receive filter design framework for the MIMO MRC system, enhancing ECM capability while ensuring radar detection and communication performance. 
	For practical implementation, the framework incorporates two distinct constraint sets: total energy (TE) and a combined TE plus peak-to-average power ratio (TE+PAPR).

	\item
	For each constraint case, we formulate a non-convex trade-off optimization problem to maximize radar SINR while minimizing communication MUI energy. 
	To integrate electronic jamming, a waveform similarity constraint on the desired jamming signal is imposed. 	
	Besides, a directional target energy constraint is adopted to focus radar energy and minimize leakage.
	Critically, we extend our framework to a discrete-sampling-based robust design to address imperfect CSI, ensuring reliable communications over a set of sampled angular uncertainties.

	\item
    To solve the formulated NP-hard problem, we develop an efficient gradient projection framework-based alternating direction method of multipliers (GP-ADMM) algorithm, which ensures fast convergence to a stationary point.
	Specifically, we treat the TE and TE+PAPR constraint sets as projection spaces and derive closed-form solutions for the optimal waveform via orthogonal projection and the bisection method (BM). The corresponding receive filter is optimally designed via a minimum variance distortionless response (MVDR) approach.

    \item
    We analyze the computational complexity of the proposed GP-ADMM algorithm and provide a theoretical proof of its convergence to a Karush-Kuhn-Tucker (KKT) point. Extensive simulation results confirm the superior capability of the MIMO MRC system, showing significant improvements in simultaneous multi-target detection, multi-user communications, and effective electronic jamming.

	\end{itemize}

    The rest of this paper is organized as follows. Section \ref{section2} describes the system model of the MIMO MRC system. In Section \ref{section3}, the optimization problems for the joint transmit waveform and receive filter design are formulated for the different cases of the considered constraints. In Section \ref{section4}, we first derive the closed-form solutions of the optimal transmit waveform and receive filter based on the proposed GP-ADMM algorithm. Then, the computational complexity and convergence of the algorithm are analyzed. Section \ref{section5} presents the simulation results and Section \ref{section6} concludes this work.\par
   
    \emph{Notations}: 
    Matrices are denoted by bold uppercase letters and vectors are denoted by bold lowercase letters. $ \mathbf{I} $  and $ \mathbf{O} $ respectively denote the identity matrix and the zero matrix, with the size determined either by a subscript or from the context. $\mathbb{C} $ and $\mathbb{R} $ are the complex and the real number domains, respectively. Superscripts $ (\cdot)^{*} $, $ (\cdot)^{T} $, $ (\cdot)^{H} $ denote the complex conjugate, transpose, and conjugate transpose, respectively. $ \mathbf{A}^{-1} $ denotes the inverse matrix of matrix $ \mathbf{A} $. $ |x| $ and $ \parallel\mathbf{x}\parallel_{2} $ represent the modulus of $ x $ and the $ \ell_{2} $ norm of the column vector $ \mathbf{x} $, respectively. $ \Re\{\cdot\}$ and $ \arg(\cdot) $  indicate the real part and the phase of a complex number, respectively. $ \text{vec}(\cdot) $ represents the vectorization operator.
	$ \nabla_{\mathbf{x}}f $ denotes the gradient of the function $ f $ with respect to the vector $ \mathbf{x} $. $ \otimes $ represents the Kronecker product.
	$ \mathbb{E} $ denotes the statistical expectation.
	\section{System Model}\label{section2}
\begin{figure}[h]
		\centering
		\includegraphics[width=1\linewidth]{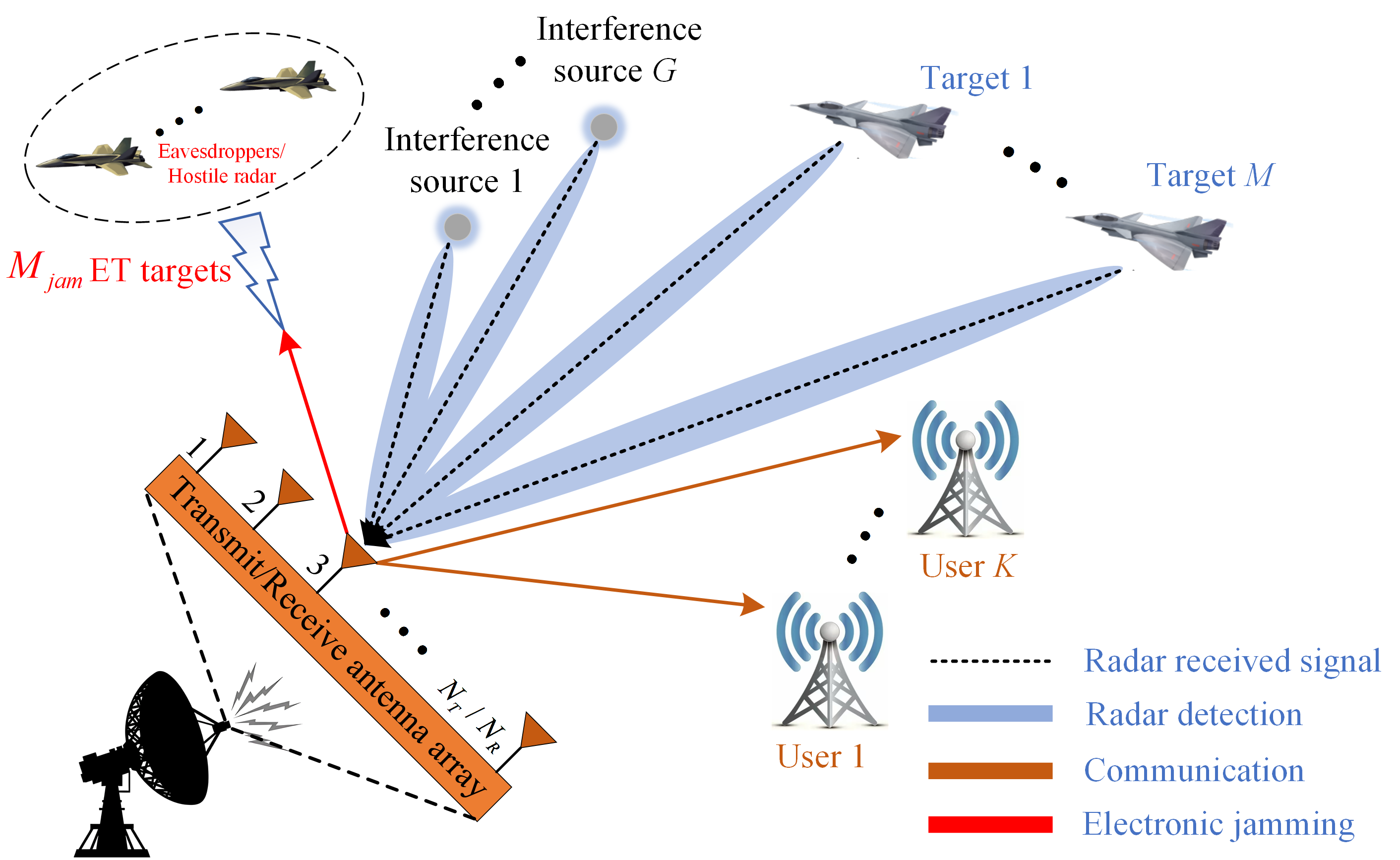}
		\caption{Illustration of a MIMO MRC system.}
		\label{fig:system_model}
	\end{figure}

	We consider a MIMO MRC system with a uniform linear array (ULA) equipped with $ N_{T} $ transmit antennas and $ N_{R} $ receive antennas, where the inter-antenna spacing is set to half-wavelength.
	As shown in Fig. \ref{fig:system_model}, the MIMO MRC system serves $ K $ single-antenna
	users in the downlink, and detects $ M $ targets while simultaneously jamming $ M_{\text{jam}} $ ET targets\footnote{The ET targets considered in this paper may be eavesdroppers or hostile radar, which can intentionally intercept the information of interest and continuously monitoring and collecting environmental information.}.
	Besides, $ G $ signal-dependent interference sources generated by surroundings exist in the system.
    Therefore, the MIMO MRC system proposed in this paper aims to communicate with friendly forces and jam ET targets. Simultaneously, the system must detect multiple targets and estimate their parameters in the presence of interference sources.
	
	Let us assume that the discrete-time baseband signal transmitted by the MIMO MRC system is $\mathbf{X}=[\mathbf{x}_{1},\mathbf{x}_{2},\cdots,\mathbf{x}_{N}]\in\mathbb{C}^{N_{T}\times N}$, where $ \mathbf{x}_{n} $ denotes the transmit vector at the $ n $-th symbol time. Then, utilizing the waveform diversity of MIMO systems, it is possible to transmit different signals in given directions, i.e., the transmitted signal at $ \theta $ is given by 	$ \mathbf{a}_{t}^{H}(\theta)\mathbf{X} $,
	where $ \mathbf{a}_{t}(\theta)=\left[1,e^{j\pi \sin(\theta)},\cdots,e^{j\pi(N_{T}-1) \sin(\theta)}\right]^{T}\in\mathbb{C}^{N_{T}\times 1}  $ denotes the steering vector of the transmit antenna array. 
	Similarly, the receive steering vector associated with an incoming signal from angle $\theta$ is defined as $ \mathbf{b}_{r}(\theta)=\left[1,e^{j\pi \sin(\theta)},\cdots,e^{j\pi(N_{R}-1) \sin(\theta)}\right]^{T}\in \mathbb{C}^{N_{R}\times 1} $.
	
	\subsection{Jamming Model}
	Let $ \bm{\Theta}_{\text{jam}}=\{\theta_{\text{jam},1},\theta_{\text{jam},2},\cdots,\theta_{\text{jam},
	M_{\text{jam}}}\} $ denote the set of directions of ET targets. Then, the signals transmitted toward the ET targets can be written as $ \mathbf{A}^{H}(\bm{\Theta}_{\text{jam}})\mathbf{X}, $
	where $\mathbf{A}(\bm{\Theta}_{\text{jam}})=[\mathbf{a}_{t}(\theta_{\text{jam},1}),\mathbf{a}_{t}(\theta_{\text{jam},2}),\cdots,\mathbf{a}_{t}(\theta_{\text{jam},M_{\text{jam}}})]\in \mathbb{C}^{N_{T}\times M_{\text{jam}}}$. Thus, to jam ET targets, $ \mathbf{X} $ should satisfy \cite{B. Tang-2022}
	\begin{equation}\label{key}
		\mathbf{A}^{H}(\bm{\Theta}_{\text{jam}})\mathbf{X}\approx \mathbf{S}_{\text{jam}},
	\end{equation}
	where $ \mathbf{S}_{\text{jam}}=[\mathbf{s}_{\text{jam},1},\mathbf{s}_{\text{jam},2},\cdots,\mathbf{s}_{\text{jam},
		M_{\text{jam}}}]^{T}\in \mathbb{C}^{M_{\text{jam}}\times N} $, and $ \mathbf{s}_{\text{jam},p} $ denotes the jamming signal expected to be emitted against the $ p $-th ET target, $ p=1,2,\cdots,M_{\text{jam}} $.
	Further, to control the transmission of different levels of jamming to ET targets in different directions, we impose the following constraints
	\begin{equation}\label{jamming constraint}
		\parallel \mathbf{G}^{H}(\theta_{\text{jam},p})\mathbf{x}- \mathbf{s}_{\text{jam},p} \parallel_{2}^{2} \leq \varepsilon_{p},p=1,2,\cdots,M_{\text{jam}},
	\end{equation}
	where $ \mathbf{G}(\theta_{\text{jam},p})=\mathbf{I}_{N}\otimes \mathbf{a}_{t}(\theta_{\text{jam},p}) $, $ \mathbf{x}=\text{vec}(\mathbf{X})=[x_{1},x_{2},\cdots,x_{N_TN}]^{T} $, and $ \varepsilon_{p} $ is a user-defined upper bound on the matching error between $ \mathbf{G}(\theta_{\text{jam},p})^{H}\mathbf{x} $ and $ \mathbf{s}_{\text{jam},p} $, where $ p=1,2,\cdots,M_{\text{jam}} $.\par
	
	\begin{Remark}\label{remark1}
	Regarding the ECM capability, it is usually determined by the energy of the jamming signals, i.e., the energy transmitted toward the ET targets. Therefore, we can use $\parallel \mathbf{G}^{H}(\theta_{\text{jam},p})\mathbf{x} \parallel_{2}^{2}$ to measure the ECM capability of the MIMO MRC system, $p=1,2,\cdots,M_{\text{jam}} $.\par
	
	Let $ \mathbf{y}_{\text{jam},p}=\mathbf{G}^{H}(\theta_{\text{jam},p})\mathbf{x} $ be the transmitted jamming signal in the direction $ \theta_{\text{jam},p} $. Note that
	\begin{equation}\label{key}
		(\parallel \mathbf{y}_{\text{jam},p} \parallel_{2}-\parallel \mathbf{s}_{\text{jam},p} \parallel_{2})^{2}\leq \parallel \mathbf{y}_{\text{jam},p}- \mathbf{s}_{\text{jam},p} \parallel_{2}^{2} \leq \varepsilon_{p}.
	\end{equation}
    Then, the jamming power at $ \theta_{\text{jam},p} $ satisfies (assuming that $\parallel \mathbf{s}_{\text{jam},p} \parallel_{2}^{2}\gg \varepsilon_{p}  $)
    \begin{equation}\label{key}
    	(\parallel \mathbf{s}_{\text{jam},p} \parallel_{2}-\sqrt{\varepsilon_{p}})^{2}\leq \parallel \mathbf{y}_{\text{jam},p} \parallel_{2}^{2} \leq (\parallel \mathbf{s}_{\text{jam},p} \parallel_{2}+\sqrt{\varepsilon_{p}})^{2}.
    \end{equation}
    Thus, the jamming power is determined by the desired transmitted jamming signal power $ P_{\text{jam},p}=\parallel \mathbf{s}_{\text{jam},p} \parallel_{2}^{2} $ and the threshold value $ \varepsilon_{p} $. The smaller $ \varepsilon_{p} $, the closer the jamming power is to the desired transmitted jamming signal power.
    
	\end{Remark}

    \subsection{Radar Model}
	A MIMO system can be configured in various modes, offering higher spatial degrees of freedom and enabling simultaneous multi-beam synthesis toward radar targets \cite{X. Liu-2020}. In this paper, we assume that $ M $ targets are located at the azimuth angles $ \theta_{m} $
	and range bins $ l_{m}, m=1,2,\cdots,M  $, and $ l_{1}\leq l_{2}\leq \cdots \leq l_{M} $.
	We also assume that $ G $ signal-dependent interference sources located at the azimuth angles $\phi_{g}\neq \theta_{m} $ and range bins $ \bar{l}_{g}, g=1,2,\cdots,G $, where $ \bar{l}_{g}\in\{l_{1}-(N-1),l_{1}-(N-2),\cdots,l_{M}+N-1)\} $. However, $ \theta_{m} $ and $ \phi_{g} $ cannot always be accurately estimated due to the steering mismatches \cite{A. Aubry-2016}, we model them as independent random variables that are uniformly distributed in $ \theta_{m}\sim U(\bar{\theta}_{m}-\vartheta_{m}/2,\bar{\theta}_{m}+\vartheta_{m}/2), \phi_{g}\sim U(\bar{\phi}_{g}-\varphi_{g}/2,\bar{\phi}_{g}+\varphi_{g}/2)$, where $ \bar{\theta}_{m} $ and $ \bar{\phi}_{g} $ denote the mean azimuths, and $ \vartheta_{m} $ and $ \varphi_{g} $ denote the azimuth errors \cite{X. Yu-2020}.\par
	
	We further assume that the self-interference between the transmit and the radar receive arrays can be reduced by employing an appropriate method \cite{A. Sabharwal-2014}.  Then, the radar signal received by the MIMO MRC system can be written as
	\begin{equation}\label{radar echo}
		\mathbf{Y}_{t}=\sum_{m=1}^{M}\alpha_{m}\mathbf{A}(\theta_{m})\tilde{\mathbf{X}}\mathbf{J}_{l_{m}-l_{1}}+\sum_{g=1}^{G}\bar{\alpha}_{g}\mathbf{A}(\phi_{g})\tilde{\mathbf{X}}\mathbf{J}_{\bar{l}_{g}-l_{1}}+\mathbf{N},
	\end{equation}
    where $ \alpha_{m} $ and $ \bar{\alpha}_{g} $ denote the amplitude gain of the $ m $-th target and the $ g $-th interference source with $ \mathbb{E}\{|\alpha_{m}|^{2}\}=\sigma_{m}^{2}$, and $\mathbb{E}\{|\bar{\alpha}_{g}|^{2}\}=\bar{\sigma}_{g}^{2} $, respectively; $ \mathbf{A}(\theta)=\mathbf{b}_{r}(\theta)\mathbf{a}_{t}^{H}(\theta)\in\mathbb{C}^{N_{R}\times N_{T}} $,  
    $ \tilde{\mathbf{X}}=[\mathbf{X},\mathbf{O}_{N_{T}\times (l_{M}-l_{1})}]\in \mathbb{C}^{N_{T}\times (N+l_{M}-l_{1})}$, $ \mathbf{J}_{l}\in\mathbb{R}^{(N+l_{M}-l_{1})\times (N+l_{M}-l_{1})} $ is a shift matrix, which is given by
    \begin{equation}\label{key}
    	\mathbf{J}_{l}(i,j)=\begin{cases}
    		1,\,\text{if}\, j-i=l,\\
    		0,\, \text{otherwise},
    	\end{cases}
    \end{equation}
    and $ \mathbf{N}\in\mathbb{C}^{N_{R}\times (N+l_{M}-l_{1})}$ denotes an additive noise matrix. \par

	By setting $ \mathbf{y}_{t}=\text{vec}(\mathbf{Y}_{t}) $ and $ \mathbf{n}=\text{vec}(\mathbf{N}) $, \eqref{radar echo} can be rewritten as
	\begin{equation}\label{key}
		\mathbf{y}_{t}=\sum_{m=1}^{M}\alpha_{m}\mathbf{\Lambda}(\theta_{m},l_{m})\mathbf{x}+\sum_{g=1}^{G}\bar{\alpha}_{g}\mathbf{\Lambda}(\phi_{g},\bar{l}_{g})\mathbf{x}+\mathbf{n},
	\end{equation}
	where $ \mathbf{\Lambda}(\theta,l) $ is determined by the azimuth angle $ \theta $ and range bin $ l $, given by
	\begin{equation}\label{key}
		\mathbf{\Lambda}(\theta,l)=(\mathbf{J}_{l-l_{1}}^{T}\otimes \mathbf{A}(\theta))[\mathbf{I}_{N_{T}N},\mathbf{O}_{N_{T}N\times N_{T}(l_{M}-l_{1})}]^{T},
	\end{equation}
     and $ \mathbf{n}\sim \mathcal{CN}(0,\sigma_{n}^{2}\mathbf{I}_{N_{R}(N+l_{M}-l_{1})})  $ with variance $\sigma_{n}^{2} $.

	The detection probability of the radar target is usually a monotonically increasing function of the SINR for the case of Gaussian interference \cite{G. Cui-2014}. Thus, we design the optimal waveform by maximizing the output SINR for the MIMO MRC system. Specifically, we use a linear finite impulse response (FIR) filter $ \mathbf{w}_{m}\in\mathbb{C}^{ N_{R}(N+l_{M}-l_{1})\times 1},m=1,2,\cdots,M $, at the radar receiver to maximize the output SINR of the $ m $-th target. In this way, the filter output of the $ m $-th target is given by
	\begin{equation}\label{key}
		\begin{split}
			r_{m}=&\mathbf{w}_{m}^{H}\mathbf{y}_{t}\\
			=&\alpha_{m}\mathbf{w}_{m}^{H}\mathbf{\Lambda}(\theta_{m},l_{m})\mathbf{x}+\mathbf{w}_{m}^{H}\sum_{i=1,i\neq m}^{M}\alpha_{i}\mathbf{\Lambda}(\theta_{i},l_{i})\mathbf{x}\\
		    &+\mathbf{w}_{m}^{H}\sum_{g=1}^{G}\bar{\alpha}_{g}\mathbf{\Lambda}(\phi_{g},\bar{l}_{g})\mathbf{x}+\mathbf{w}_{m}^{H}\mathbf{n}.
		\end{split}
	\end{equation}
	 For simplicity, we let $ \theta_{i} $, $ l_{i} $ and $\sigma_{i},i=1,2\cdots,M+G $, denote the $ i $-th elements of $ (\theta_{1},\cdots,\theta_{M},\phi_{1},\cdots,\phi_{G}) $, $ (l_{1},\cdots,l_{M},\bar{l}_{1},\cdots,\bar{l}_{G}) $ and $ (\sigma_{1},\cdots,\sigma_{M},\bar{\sigma}_{1},\cdots,\bar{\sigma}_{G}) $ respectively, and set $ Q=N_{R}(N+l_{M}-l_{1}) $. Then, the output $ \text{SINR}_{m} $ of the $ m $-th target can be expressed as
	\begin{equation}\label{key}
		\begin{split}
			&\text{SINR}_{m}\\
			=&\dfrac{\mathbb{E}[|\alpha_{m}\mathbf{w}_{m}^{H}\mathbf{\Lambda}(\theta_{m},l_{m})\mathbf{x}|^{2}]}{\mathbb{E}[|\mathbf{w}_{m}^{H}(\sum\limits_{i=1,i\neq m}^{M}\alpha_{i}\mathbf{\Lambda}(\theta_{i},l_{i})\mathbf{x}+\sum\limits_{g=1}^{G}\bar{\alpha}_{g}\mathbf{\Lambda}(\phi_{g},\bar{l}_{g})\mathbf{x}+\mathbf{n})|^{2}]}\\
			=&\dfrac{\sigma_{m}^{2}|\mathbf{w}_{m}^{H}\mathbf{\Lambda}(\theta_{m},l_{m})\mathbf{x}|^{2}}{\mathbf{w}_{m}^{H}(\bm{\Sigma}_{m}(\mathbf{x})+\sigma_{n}^{2}\mathbf{I}_{Q})\mathbf{w}_{m}},
		\end{split}
	\end{equation}
	where $ \bm{\Sigma}_{m}(\mathbf{x})=\sum_{i=1,i\neq m}^{M+G}\sigma_{i}^{2}\mathbf{\Lambda}(\theta_{i},l_{i})\mathbf{x}\mathbf{x}^{H}\mathbf{\Lambda}(\theta_{i},l_{i})^{H}$. Thus, the optimal waveform $ \mathbf{x} $ and filters $ \{\mathbf{w}_{m}\}_{m=1}^{M} $ can be derived via the SINR maximization criterion. 
	\begin{Remark}\label{remark2}
		We assume that the MIMO MRC system can obtain pestimates of the locations of ET targets in the previous stages \cite{B. Tang-2022}. To update the information about $ \bm{\Theta}_{\text{jam}} $,  the system can transmit detection waveforms and estimate $ \bm{\Theta}_{\text{jam}} $ from the echoes \cite{L. Xu-2008}. Moreover, prior knowledge about targets and interference can be obtained via a potentially cognitive paradigm, which can utilize an environmental dynamic database to obtain the azimuth angles $ \theta_{i} $  and range bins $ l_{i}, i=1,2,\cdots,M+G $ \cite{A. Aubry-2013}. However, from a practical point of view, the exact knowledge of $ \theta_{i} $ is not always available due to the steering mismatches \cite{A. Aubry-2016}. Thus, the average of multiple angle observations is employed as a deterministic input for waveform design.
	\end{Remark}\par
	To minimize energy leakage and control energy distribution among multiple targets, a key approach is to maximize the transmitted energy toward the intended target directions. To this end, we impose the following constraints
	\begin{equation}\label{target energy constraint}
		\parallel\mathbf{T}_{m}^{H}\mathbf{x}\parallel_{2}^{2}\geq e_{m},m=1,2,\cdots,M,
	\end{equation}
	where $ \mathbf{T}_{m}=\mathbf{I}_{N}\otimes \mathbf{a}_{t}(\theta_{m}) $, and $ e_{m} $ is the minimum allowed energy of the signal transmitted toward the $ m $-th detection target, $ m=1,2,\cdots,M $.

	\subsection{Communication Model}
	In addition to its jamming and radar sensing capabilities, the MIMO MRC system simultaneously delivers information symbols to $ K $ single-antenna users using the same transmit waveform. However, a key challenge arises when the ET targets are located close to the communication users in the angular domain, resulting in strong spatial correlation between their corresponding array responses. Moreover, the system-generated jamming signals may reach the communication users through multipath propagation, potentially causing additional interference. To characterize these propagation effects, we adopt a Rician fading channel model for the downlink communication links. Accordingly, the received signal by the downlink users over $ N $ symbol intervals is given by
	
	\begin{equation}\label{key}
		\mathbf{Y}=\mathbf{HX}+\mathbf{Z},
	\end{equation}
	where $ \mathbf{Y}\in\mathbb{C}^{K\times N} $ is a matrix whose entry $ y_{k,n} $ is the signal
	received by the $ k $-th user at the $ n $-th symbol time, $ 1\leq k\leq K$, and $1\leq n\leq N$. $ \mathbf{H}=[\mathbf{h}_{1},\mathbf{h}_{2},\cdots,\mathbf{h}_{K}]^{T}\in\mathbb{C}^{K\times N_{T}} $ is the Rician fading channel matrix between the MIMO MRC system and users, 
	and $ \mathbf{Z}=[\mathbf{z}_{1},\mathbf{z}_{2},\cdots,\mathbf{z}_{N}] $ is the additive white Gaussian noise (AWGN) matrix, where $ \mathbf{z}_{n}\sim \mathcal{CN}(0,\sigma_{z}^{2}\mathbf{I}_{K})$ and $ \sigma_{z}^{2} $ denotes the noise variance.\par
	
	Specifically, $ \mathbf{h}_{k}\in\mathbb{C}^{N_{T}\times 1}$ denotes the channel coefficients between the system transmitter and the $ k $-th user, which can be efficiently represented as
	\begin{equation}\label{key}
		\mathbf{h}_{k}=\sqrt{\dfrac{K_{a}}{K_{a}+1}}\bar{\mathbf{h}}_{k}+\sqrt{\dfrac{1}{K_{a}+1}}\tilde{\mathbf{h}}_{k},
	\end{equation}
	where $ K_{a} $ is the Rician factor, $ \bar{\mathbf{h}}_{k}=\mathbf{a}_{t}(\theta_{c,k}) $ represents the line-of-sight (LoS) component and $ \theta_{c,k} $ is the azimuth of the $ k $-th user,  and $ \tilde{\mathbf{h}}_{k}\sim \mathcal{CN}(0,\mathbf{I}_{N_{T}})  $ is the non-LoS (NLoS) component, which is
	assumed to be perfectly estimated. \par

	Furthermore, we assume that in the $ n $-th symbol time, the $ k $-th user expects to receive a communication data symbol $ s_{k,n}\in\mathcal{O},1\leq k\leq K, 1\leq n\leq N $, where $ \mathcal{O} $ is the set of the employed constellation points and $ \mathbf{S}=[\mathbf{s}_{1},\mathbf{s}_{2},\cdots,\mathbf{s}_{N}]\in \mathbb{C}^{K\times N} $ is the desired constellation symbol matrix for the downlink users. Then, the communication part of the MIMO MRC system aims at the design of a transmit signal vector $ \mathbf{x}_{n} $, such that the received vector signal $ \mathbf{y}_{n} $ is as close to $ \mathbf{s}_{n} $, as possible. Thus, the received signal of the $ k $-th user in the $ n $-th symbol time can be written as
	\begin{equation}\label{key}
		y_{k,n}=s_{k,n}+\underbrace{(\mathbf{h}_{k}^{T}\mathbf{x}_{n}-s_{k,n})}_{\text{MUI at the $ k $-th user} }+z_{k,n},
	\end{equation}
	where  $ \mathbf{h}_{k}^{T}\mathbf{x}_{n}-s_{k,n} $ denotes the MUI at the $ k $-th user. For this, the receive SINR per block of $ N  $ symbols for the $ k $-th user is defined as
	\begin{equation}\label{key}
		\gamma_{k}=\dfrac{\mathbb{E}\{|s_{k,n}|^{2}\}}{\mathbb{E}\{|\mathbf{h}_{k}^{T}\mathbf{x}_{n}-s_{k,n}|^{2}\}+\sigma_{z}^{2}},
	\end{equation}
	where $ \mathbb{E} $ denotes the ensemble average with respect to the time index. It follows that the
	achievable sum-rate (ASR) of the users can be given as
	\begin{equation}\label{ASR}
		R=\sum_{k=1}^{K}\log_{2}(1+\gamma_{k}).
	\end{equation}\par
   In the following, we give the communication performance optimization criterions for the case of perfect CSI and imperfect CSI respectively.

	\begin{Remark}\label{remark3}
		Since the transmitted communication symbols are drawn from a finite constellation and the symbol-level waveform mismatch is not necessarily Gaussian, \eqref{ASR} is used as a rate surrogate measure rather than an exact ASR expression in this paper \cite{F. Liu-2018-2}. In the subsequent simulations, ASR denotes a surrogate measure for comparing the communication performance under different design settings.		
	\end{Remark}\par

	\subsubsection{Communication criterion with perfect CSI}

	For a given constellation with fixed energy, the power of the desired signal $ \mathbb{E}\{|s_{k,n}|^{2}\} $ is also fixed. Therefore, we can maximize the communication SINR by minimizing the MUI energy $ \mathbb{E}\{|\mathbf{h}_{k}^{T}\mathbf{x}_{n}-s_{k,n}|^{2}\} $, which maximizes the communication ASR of the users. Further, the MUI energy of $ K $ users can be rewritten as
	\begin{equation}\label{key}
		\bm{\Xi}\triangleq\parallel \tilde{\mathbf{H}}\mathbf{x}-\mathbf{s}\parallel_{2}^{2},
	\end{equation}
	where $ \tilde{\mathbf{H}}=\mathbf{I}_{N}\otimes \mathbf{H} $, and $ \mathbf{s}=\text{vec}(\mathbf{S})$.
    
	\subsubsection{Communication criterion with imperfect CSI}
	It is worth noting that in practice, the transmitter typically cannot always obtain perfect communication CSI.
	In the following, we will focus on the imperfect communication CSI.\par
    
    Unlike the bounded channel error model \cite{S. Wang-2024}, we consider the case that the angle
	uncertainty interval of the $ k $-th user is roughly known within the angle interval $ [\theta_{c,k}-\Delta\theta_{c},\theta_{c,k}+\Delta\theta_{c}] $. Further, we discretize the azimuth interval of the $ k $-th user into $ D $ angles, where $ \theta_{c,k}^{(q)}= \theta_{c,k}-\Delta\theta_{c}+2(q-1)\Delta\theta_{c}/(D-1),q=1,2,\cdots,D$. Then, the channel between the system transmitter and the $ k $-th user located at angle $ \theta_{c,k}^{(q)} $ can be represented as
	\begin{equation}\label{key}
		\mathbf{h}_{k}^{(q)}=\sqrt{\dfrac{K_{a}}{K_{a}+1}}\mathbf{a}_{t}(\theta_{c,k}^{(q)})+\sqrt{\dfrac{1}{K_{a}+1}}\tilde{\mathbf{h}}_{k},
	\end{equation}
	and $ \mathbf{H}^{(q)}=[\mathbf{h}_{1}^{(q)},\mathbf{h}_{2}^{(q)},\cdots,\mathbf{h}_{K}^{(q)}]^{T} $ is defined accordingly.\par
	
	To realize communication information transmission under imperfect CSI, we minimize the MUI energy of $ K $ users in the angle uncertainty interval $ [\theta_{c,k}-\Delta\theta_{c},\theta_{c,k}+\Delta\theta_{c}]$, where $k=1,2,\cdots,K $. In this way, the communication optimization criterion can be given as
	\begin{equation}\label{key}
		\bm{\Xi}\triangleq\sum_{q=1}^{D}\omega_{q}\parallel \tilde{\mathbf{H}}^{(q)}\mathbf{x}-\mathbf{s}\parallel_{2}^{2},
	\end{equation}
	where  $ \omega_{q}$, $q=1,2,\cdots,D $ are
	weight factors that control the communication performance of different angles, $ \tilde{\mathbf{H}}^{(q)}=\mathbf{I}_{N}\otimes \mathbf{H}^{(q)}$, and $ \mathbf{s}=\text{vec}(\mathbf{S})$.

	\section{Problem Formulation}\label{section3}
	In this section, the cases of TE and TE+PAPR constraints are considered, respectively. For each case, we formulate an optimization problem by jointly designing the optimal transmit waveform and the radar receive filter. In the following, let us focus on how the functions of radar, communication, and jamming can be simultaneously optimized.\par
	
	The communication part of the MIMO MRC system requires that the MUI energy is minimized to achieve the transmission of the desired symbol information to the intended user. The radar part of the system can maximize the detection probability by maximizing the SINR for $ M $ targets. In addition, the directional target energy constraints in\eqref{target energy constraint} are adopted to minimize radar transmitter energy leakage in target directions.
	For the purposes of jamming, we further consider the inequality constraints in\eqref{jamming constraint} to transmit jamming of different levels to ET targets. It is easy to see that the simultaneous optimization of both of the communication and radar functions under the jamming functionality constraint is not a straightforward task. Therefore, we opt for the weighted sum method that combines the communication MUI energy and the terms $ 1/\text{SINR}_{m}$, $m =1,2,\cdots,M $ into a composite objective. Then, the composite objective function is given by
	\begin{equation}\label{objective function}
			f(\mathbf{x},\{\mathbf{w}_{m}\}_{m=1}^{M})
			=\rho\bm{\Xi}+(1-\rho)\sum_{m=1}^{M}\dfrac{1}{\text{SINR}_{m}},
	\end{equation}
    where $ \rho\in [0,1] $ is the weighting factor that determines the trade-off between communication and radar performance in system. It is evident from\eqref{objective function} that the larger $ \rho $ imposes a greater weight on the communication performance, in which case the communication performance of the system outperforms the radar performance. In contrast, a higher weight is allocated to the minimization
    of $ 1/\text{SINR}_{m}, m =1,2,\cdots,M $ or equivalently to the the maximization of the radar SINR. In this way, a flexible performance trade-off between objectives of the communications and the radar is achieved.\par
	
	\subsection{TE Constraint}
	It is worth noting that the TE constraint is widely used in the radar domain to simulate the limits of transmitter power amplifiers. Under this constraint, the corresponding design problem for the MIMO MRC system is formulated as $\mathcal{P}_{0,1}$.
	\begin{equation}\label{key}
			\begin{aligned}
				\min_{\mathbf{x},\{\mathbf{w}_{m}\}_{m=1}^{M}}\;& f(\mathbf{x},\{\mathbf{w}_{m}\}_{m=1}^{M})&\\
				\mbox{s.t.}\;\;\;\;& \parallel \mathbf{G}^{H}(\theta_{\text{jam},p})\mathbf{x}- \mathbf{s}_{\text{jam},p} \parallel_{2}^{2} \leq \varepsilon_{p},p=1,\cdots,M_{\text{jam}},&\\
				& \parallel\mathbf{T}_{m}^{H}\mathbf{x}\parallel_{2}^{2}\geq e_{m},m=1,2,\cdots,M,&\\
				& \parallel \mathbf{x}\parallel_{2}^{2}\leq P_{T},&\\
			\end{aligned}
	\end{equation}
	where $ P_{T} $ is the maximum transmission energy supported by the MIMO MRC system.
	
	\subsection{TE+PAPR Constraint}
	In practical communication and radar systems, the transmitted waveforms with low peak-to-average power ratio (PAPR) can ensure that the power amplifiers to operate at maximum efficiency while avoiding nonlinear effect.
	Therefore, we enforce the following constraints
    \begin{equation}\label{TE+PAPR_constraint}
    	\parallel \mathbf{x}\parallel_{2}^{2}= P_{T},\;\text{PAPR}(\mathbf{x})\leq \eta,
    \end{equation}
	where $ \eta $ represents the PAPR upper bound of the waveform $ \mathbf{x} $, and the PAPR of $ \mathbf{x} $ is defined as
    \begin{equation}\label{key}
    	\text{PAPR}(\mathbf{x})=\dfrac{\max\limits_{i=1,2,\cdots,N_{T}N}\{|x_{i}|^{2}\}}{\parallel \mathbf{x} \parallel_{2}^{2}/N_{T}N}.
    \end{equation}
	For the special case of $ \eta = 1 $, the PAPR constraint in\eqref{TE+PAPR_constraint} becomes the constant-modulus (CM) constraint, i.e.,
	\begin{equation}\label{key}
		|x_{i}|=\sqrt{\dfrac{P_{T}}{N_{T}N}},i=1,2,\cdots,N_{T}N.
	\end{equation}
	Therefore, to design the low PAPR optimal transmit waveform and the radar receive filter for the MIMO MRC system, we formulate the following problem $ \mathcal{P}_{0,2} $
	\begin{equation}\label{key}
			\begin{aligned}
				\min_{\mathbf{x},\{\mathbf{w}_{m}\}_{m=1}^{M}}\;& f(\mathbf{x},\{\mathbf{w}_{m}\}_{m=1}^{M})&\\
				\mbox{s.t.}\;\;\;\;& \parallel \mathbf{G}^{H}(\theta_{\text{jam},p})\mathbf{x}- \mathbf{s}_{\text{jam},p} \parallel_{2}^{2} \leq \varepsilon_{p},p=1,\cdots,M_{\text{jam}},&\\
				& \parallel\mathbf{T}_{m}^{H}\mathbf{x}\parallel_{2}^{2}\geq e_{m},m=1,2,\cdots,M,&\\
				& \parallel \mathbf{x}\parallel_{2}^{2}= P_{T},&\\
				& \text{PAPR}(\mathbf{x})\leq \eta.&\\
			\end{aligned}
	\end{equation}\par
	Next, we consider the two different constraint cases discussed above and proceed to discuss the solution of the different problems.

	\section{Joint Design Algorithm for Optimal Transmit Waveforms and Radar Receive Filters}\label{section4}
	In this section, we develop algorithmic solutions to the optimization problems $ \mathcal{P}_{0,1} $ and $ \mathcal{P}_{0,2} $. The aforementioned problems are difficult to solve directly due to the involved non-convex objective function and the non-convex constraint in $ \mathcal{P}_{0,1} $ and $ \mathcal{P}_{0,2} $. Furthermore, we observe that all problems involve a differentiable objective function and that the constraint sets of problems  $ \mathcal{P}_{0,1} $ and $ \mathcal{P}_{0,2} $ can be obtained as closed solutions using orthogonal projection. Therefore, we next address the proposed optimization problem through the ADMM-based approach and GP framework.\par

	Before proceeding towards the algorithmic solutions development, we redefine the constraint sets of problems $ \mathcal{P}_{0,1} $ and $ \mathcal{P}_{0,2} $. To simplify the representation, we define set $ \bm{\Omega} $ to denote the constraint sets of $ \mathcal{P}_{0,1} $ and $ \mathcal{P}_{0,2} $, which can given by
	\begin{equation}\label{key}
		\bm{\Omega}_{\text{TE}}=\{\mathbf{x}\in \mathbb{C}^{N_TN\times 1} | \parallel \mathbf{x}\parallel_{2}^{2}\leq P_{T}\},
	\end{equation}
	\begin{equation}\label{key}
		\bm{\Omega}_{\text{TE+PAPR}}=\{\mathbf{x}\in \mathbb{C}^{N_TN\times 1} | \parallel \mathbf{x}\parallel_{2}^{2}= P_{T},\text{PAPR}(\mathbf{x})\leq \eta\}.
	\end{equation}\par
    
    Moreover, given that the same solution method applies to both perfect and imperfect CSI cases, we deem it redundant to analyze them separately. The distinctions between the two cases are detailed in \emph{Remark} \ref{remark4}.

	\subsection{Joint Design Algorithm based on GP-ADMM}
	
    Obviously, the problems $ \mathcal{P}_{0,1} $ and $ \mathcal{P}_{0,2} $ share the same objective function, jamming functionality and target energy constraints. To this end, we first introduce the auxiliary variable $ \{\mathbf{v}_{p}\} _{p=1}^{M_{\text{jam}}}$ such that
	\begin{equation}\label{key}
		\mathbf{v}_{p}= \mathbf{G}^{H}(\theta_{\text{jam},p})\mathbf{x}- \mathbf{s}_{\text{jam},p},\parallel\mathbf{v}_{p} \parallel_{2}^{2} \leq \varepsilon_{p},p=1,2,\cdots,M_{\text{jam}}.
	\end{equation}
    Similarly, we define $ \mathbf{g}_{m}=\mathbf{T}_{m}^{H}\mathbf{x},m=1,2,\cdots,M $, and the target energy constraints can be rewritten as 
    \begin{equation}\label{key}
    	\mathbf{g}_{m}=\mathbf{T}_{m}^{H}\mathbf{x},\parallel\mathbf{g}_{m} \parallel_{2}^{2}\geq e_{m},m=1,2,\cdots,M.
    \end{equation}
    
	Then, we can obtain the augmented Lagrange function for problems $ \mathcal{P}_{0,1} $ and $ \mathcal{P}_{0,2} $ in the ADMM framework
	\begin{equation}\label{Lagrange}
		\begin{split}
			&\mathcal{L}_{\lambda}(\mathbf{x},\{\mathbf{w}_{m}\}_{m=1}^{M},\{\mathbf{v}_{p}\} _{p=1}^{M_{\text{jam}}},\{\mathbf{g}_{m}\}_{m=1}^{M},\{\mathbf{y}_{p}\} _{p=1}^{M_{\text{jam}}},\{\mathbf{t}_{m}\}_{m=1}^{M})\\
			&=\rho\bm{\Xi}
			+(1-\rho)\sum_{m=1}^{M}\dfrac{1}{\text{SINR}_{m}}\\
			&+\dfrac{\lambda}{2}\sum_{p=1}^{M_{\text{jam}}}\left( \parallel\mathbf{v}_{p}-\mathbf{G}^{H}(\theta_{\text{jam},p})\mathbf{x}+\mathbf{s}_{\text{jam},p}+\mathbf{y}_{p} \parallel_{2}^{2}- \parallel\mathbf{y}_{p} \parallel_{2}^{2} \right)\\
			&+\dfrac{\lambda}{2}\sum_{m=1}^{M}\left( \parallel\mathbf{g}_{m}-\mathbf{T}_{m}^{H}\mathbf{x}+\mathbf{t}_{m} \parallel_{2}^{2}- \parallel\mathbf{t}_{m} \parallel_{2}^{2} \right),
		\end{split}
	\end{equation}
	where $ \lambda $ is a penalty parameter, and $ \{\mathbf{y}_{p}\} _{p=1}^{M_{\text{jam}}} $ and $ \{\mathbf{t}_{m}\}_{m=1}^{M} $ are the Lagrange multipliers. In the $ t+ 1 $-th iteration of the ADMM algorithm, the variables $ \mathbf{x},\{\mathbf{w}_{m}\}_{m=1}^{M},\{\mathbf{v}_{p}\}_{p=1}^{M_{\text{jam}}}, \{\mathbf{g}_{m}\}_{m=1}^{M},\{\mathbf{y}_{p}\} _{p=1}^{M_{\text{jam}}}$ and $\{\mathbf{t}_{m}\}_{m=1}^{M} $ are sequentially updated mainly by the following optimization problem (for simplicity, we omit the variable superscript):
	\begin{subequations}\label{Eq1:GP-ADMM}
		\begin{align}
				\mathbf{x}^{(t+1)}&=\arg \min_{\mathbf{x}\in \bm{\Omega}} \mathcal{L}_{\lambda}(\mathbf{x}^{(t)},\mathbf{w}_{m}^{(t)},\mathbf{v}_{p}^{(t)},\mathbf{g}_{m}^{(t)} ,\mathbf{y}_{p}^{(t)},\mathbf{t}_{m}^{(t)} ),\label{Eq1:GP-ADMM:sub1}\\
				\mathbf{w}_{m}^{(t+1)}&=\arg \min_{\mathbf{w}_{m}} \mathcal{L}_{\lambda}(\mathbf{x}^{(t+1)},\mathbf{w}_{m}^{(t)},\mathbf{v}_{p}^{(t)},\mathbf{g}_{m}^{(t)} ,\mathbf{y}_{p}^{(t)},\mathbf{t}_{m}^{(t)}),\label{Eq1:GP-ADMM:sub2}\\
				\mathbf{v}_{p}^{(t+1)}&=\arg\min_{\mathbf{v}_{p} } \mathcal{L}_{\lambda}(\mathbf{x}^{(t+1)},\mathbf{w}_{m}^{(t+1)},\mathbf{v}_{p}^{(t)},\mathbf{g}_{m}^{(t)} ,\mathbf{y}_{p}^{(t)},\mathbf{t}_{m}^{(t)} ),\label{Eq1:GP-ADMM:sub3}\\
				\mathbf{g}_{m}^{(t+1)}&=\arg\min_{\mathbf{g}_{m}} \mathcal{L}_{\lambda}(\mathbf{x}^{(t+1)},\mathbf{w}_{m}^{(t+1)},\mathbf{v}_{p}^{(t+1)},\mathbf{g}_{m}^{(t)} ,\mathbf{y}_{p}^{(t)},\mathbf{t}_{m}^{(t)} ),\label{Eq1:GP-ADMM:sub4}\\
			\mathbf{y}_{p}^{(t+1)}&=\mathbf{y}_{p}^{(t)}+\mathbf{v}_{p}^{(t+1)}-\mathbf{G}^{H}(\theta_{\text{jam},p})\mathbf{x}^{(t+1)}+\mathbf{s}_{\text{jam},p},\label{Eq1:GP-ADMM:sub5}\\
			\mathbf{t}_{m}^{(t+1)}&=\mathbf{t}_{m}^{(t)}+\mathbf{g}_{m}^{(t+1)}-\mathbf{T}_{m}^{H}\mathbf{x}^{(t+1)}.\label{Eq1:GP-ADMM:sub6}
		\end{align}
	\end{subequations}
    In the following, we present solutions to the optimization problems in\eqref{Eq1:GP-ADMM:sub1},\eqref{Eq1:GP-ADMM:sub2},\eqref{Eq1:GP-ADMM:sub3} and\eqref{Eq1:GP-ADMM:sub4}.\par
    \subsubsection{Update of transmit waveform $ \mathbf{x}$}
    Since the subproblem of updating the waveform $ \mathbf{x}$ in\eqref{Eq1:GP-ADMM:sub1} is non-convex and generally difficult to solve optimally, we employ a single GP step to obtain an approximate solution at each iteration, whose core iteration steps are given as
    
    \begin{equation}\label{GP}
    	\mathbf{d}^{(t+1)}=\mathbf{x}^{(t)}-\alpha^{(t)} \nabla_{\mathbf{x}}\mathcal{L}_{\lambda}(\mathbf{x}^{(t)},\mathbf{w}_{m}^{(t)},\mathbf{v}_{p}^{(t)},\mathbf{g}_{m}^{(t)} ,\mathbf{y}_{p}^{(t)},\mathbf{t}_{m}^{(t)}),
    \end{equation}
    \begin{equation}\label{key}
    	\mathbf{x}^{(t+1)}=\bm{\Pi}_{\bm{\Omega}}\{\mathbf{d}^{(t+1)}\},
    \end{equation}
    where $ \alpha^{(t)} $ is a suitable step-size parameter in the $ t $-th iteration, $ \bm{\Pi}_{\bm{\Omega}}\{\cdot\} $ denotes the orthogonal projection operator onto set $ \bm{\Omega} $. Depending on the problem under consideration, $ \bm{\Omega} $ can be either $ \bm{\Omega}_{\text{TE}} $ or $ \bm{\Omega}_{\text{TE+PAPR}} $. Furthermore, the gradient vectors in\eqref{GP} is given by
    \begin{equation}\label{key}
    	\begin{split}
    		&\nabla_{\mathbf{x}}\mathcal{L}_{\lambda}(\mathbf{x}^{(t)},\mathbf{w}_{m}^{(t)},\mathbf{v}_{p}^{(t)},\mathbf{g}_{m}^{(t)} ,\mathbf{y}_{p}^{(t)},\mathbf{t}_{m}^{(t)})\\
    		=&\rho\nabla_{\mathbf{x}}\bm{\Xi}\\
    		&+2(1-\rho)\sum_{m=1}^{M}\left[\dfrac{\bm{\Gamma}(\mathbf{w}_{m}^{(t)})\mathbf{x}^{(t)}}{a_{m}^{(t)}}-\dfrac{\bm{\Phi}(\mathbf{x}^{(t)},\mathbf{w}_{m}^{(t)})\bm{\Psi}(\mathbf{w}_{m}^{(t)})\mathbf{x}^{(t)}}{|a_{m}^{(t)}|^{2}}\right]\\
    		&+\dfrac{\lambda}{2}\sum_{p=1}^{M_{\text{jam}}}\left[2\mathbf{G}(\theta_{\text{jam},p})\mathbf{G}^{H}(\theta_{\text{jam},p})\mathbf{x}^{(t)}\right]\\
    		&-\dfrac{\lambda}{2}\sum_{p=1}^{M_{\text{jam}}}\left[2\mathbf{G}(\theta_{\text{jam},p})(\mathbf{v}_{p}^{(t)}+\mathbf{s}_{\text{jam},p})+ 2\mathbf{G}(\theta_{\text{jam},p})\mathbf{y}_{p}^{(t)}\right]\\
    		&+\dfrac{\lambda}{2}\sum_{m=1}^{M}\left[2\mathbf{T}_{m}\mathbf{T}_{m}^{H}\mathbf{x}^{(t)}-2\mathbf{T}_{m}(\mathbf{g}_{m}^{(t)}+\mathbf{t}_{m}^{(t)})\right],
    	\end{split}
    \end{equation}
    where $ a_{m}^{(t)}=\sigma_{m}^{2}|(\mathbf{x}^{(t)})^{H}\mathbf{\Lambda}^{H}(\theta_{m},l_{m})\mathbf{w}_{m}^{(t)}|^{2} $, and
    \begin{equation}\label{key}
    	\bm{\Gamma}(\mathbf{w}_{m}^{(t)})=\sum_{i=1,i\neq m}^{M+G}\sigma_{i}^{2}\mathbf{\Lambda}^{H}(\theta_{i},l_{i})\mathbf{w}_{m}^{(t)}(\mathbf{w}_{m}^{(t)})^{H}\mathbf{\Lambda}(\theta_{i},l_{i}),
    \end{equation}
    \begin{equation}\label{key}
    	\bm{\Phi}(\mathbf{x}^{(t)},\mathbf{w}_{m}^{(t)})=(\mathbf{x}^{(t)})^{H}
    	\bm{\Gamma}(\mathbf{w}_{m}^{(t)})\mathbf{x}^{(t)}+\sigma_{n}^{2}(\mathbf{w}_{m}^{(t)})^{H}\mathbf{w}_{m}^{(t)},
    \end{equation}
    \begin{equation}\label{key}
    	\bm{\Psi}(\mathbf{w}_{m}^{(t)})=\sigma_{m}^{2}\mathbf{\Lambda}^{H}(\theta_{m},l_{m})\mathbf{w}_{m}^{(t)}(\mathbf{w}_{m}^{(t)})^{H}\mathbf{\Lambda}(\theta_{m},l_{m}).
    \end{equation}
    \begin{Remark}\label{remark4}
    	Note that $ \nabla_{\mathbf{x}}\bm{\Xi} $ has different gradient vectors under different CSI conditions. For perfect CSI,
    	\begin{equation}\label{key}
    		\nabla_{\mathbf{x}}\bm{\Xi}=2(\tilde{\mathbf{H}}^{H}\tilde{\mathbf{H}}\mathbf{x}^{(t)}-\tilde{\mathbf{H}}^{H}\mathbf{s}).
    	\end{equation}
        Similarly, for imperfect CSI,
    	\begin{equation}\label{key}
    		\nabla_{\mathbf{x}}\bm{\Xi}=2\sum_{q=1}^{D}\omega_{q}\left[(\tilde{\mathbf{H}}^{(q)})^{H}\tilde{\mathbf{H}}^{(q)}\mathbf{x}^{(t)}-(\tilde{\mathbf{H}}^{(q)})^{H}\mathbf{s}\right].
    	\end{equation}
    	
    \end{Remark}

    Let us now proceed to derive the projection operator $ \bm{\Pi}_{\bm{\Omega}}\{\cdot\} $. The constraint associated with the set $ \bm{\Omega}_{\text{TE}} $ can be achieved by solving the following optimization problem
    \begin{equation}\label{TE_projection}
    	\begin{aligned}
    		\min_{\mathbf{x}^{\dagger}}\;& \parallel \mathbf{x}^{\dagger}-\mathbf{d}\parallel_{2}^{2}&\\
    		\mbox{s.t.}\;& \mathbf{x}^{\dagger}\in\bm{\Omega}_{\text{TE}},& 
    	\end{aligned}
    \end{equation}
    where $ \mathbf{x}^{\dagger} $ is the projection of $ \mathbf{d} $ onto the set $ \bm{\Omega}_{\text{TE}} $. In this way, we have 
    \begin{equation}\label{key}
    	\mathbf{x}^{\dagger}=\begin{cases}
    		\mathbf{d},\;&\parallel \mathbf{d}\parallel_{2}^{2}\leq P_{T},\\
    		\sqrt{P_{T}}\frac{\mathbf{d}}{\parallel \mathbf{d}\parallel_{2}},\;&\parallel \mathbf{d}\parallel_{2}^{2}> P_{T}.
    	\end{cases}
    \end{equation}
    Similarly, the projection on the set $ \bm{\Omega}_{\text{TE+PAPR}} $ can be derived by solving a problem similar to\eqref{TE_projection}, i.e.,
    \begin{equation}\label{TE+PAPR}
    	\begin{aligned}
    		\min_{\mathbf{x}^{\ddagger}}\;& \parallel \mathbf{x}^{\ddagger}-\mathbf{d}\parallel_{2}^{2}&\\
    		\mbox{s.t.}\;& \mathbf{x}^{\ddagger}\in\bm{\Omega}_{\text{TE}+\text{PAPR}}, & 
    	\end{aligned}
    \end{equation}
     where $ \mathbf{x}^{\ddagger} $ is the projection of $ \mathbf{d} $ onto the set $ \bm{\Omega}_{\text{TE}+\text{PAPR}}$. This kind of problem is a linear optimization problem with multiple  quadratic constraints \cite{J. A. Tropp-2005}.
     Assuming that the number of nonzero elements of $ \mathbf{d} $ is $ b $, then the solution of the optimization problem\eqref{TE+PAPR} can be divided into two cases
    \begin{itemize}
    	\item If $ b\eta\leq N_{T}N $,
        \begin{equation}\label{key}
        	x^{\ddagger}_{i}=\begin{cases}
        		\delta e^{j\arg(d_{i})},&i=1,2,\cdots,b,\\
        		\sqrt{\frac{P_{T}-b\delta^{2}}{N_{T}N-b}}e^{j\arg(d_{i})},&i=b+1,\cdots,N_{T}N,
        	\end{cases}
        \end{equation}
    where $ \delta=\sqrt{\eta P_{T}/N_{T}N} $, and $ d_{i} $ is the $ i $-th element of the vector $ \mathbf{d} $.
        \item Otherwise,
    \begin{equation}\label{key}
    	x^{\ddagger}_{i}=\min\{\kappa|d_{i}|e^{j\arg(d_{i})}, \delta e^{j\arg(d_{i})}\},
    \end{equation}
    where $ \kappa\in [0,\delta/\min\{|d_{i}|,|d_{i}|\neq0\}] $ can be obtained by
    using the BM to solve the following equation \cite{Y. Zhou-2025}
    \begin{equation}\label{key}
    	\sum_{i=1}^{N_{T}N}\min\{\kappa^{2}|d_{i}|^{2},\delta^{2}\}=P_{T}.
    \end{equation}
    \end{itemize}\par
    \subsubsection{Update of receive filter $ \mathbf{w}_{m}$}
    For the update of receive filters $ \{\mathbf{w}_{m}\}_{m=1}^{M}$, we can ignore the independent term of $ \{\mathbf{w}_{m}\}_{m=1}^{M}$ in\eqref{Eq1:GP-ADMM:sub2}, then the optimization problem\eqref{Eq1:GP-ADMM:sub2} can be simplified as 
    \begin{equation}\label{MVDR}
    	\begin{aligned}
    		\min_{\mathbf{w}_{m}}\;& (1-\rho)\dfrac{\mathbf{w}_{m}^{H}(\bm{\Sigma}_{m}(\mathbf{x}^{(t+1)})+\sigma_{n}^{2}\mathbf{I}_{Q})\mathbf{w}_{m}}{\sigma_{m}^{2}|\mathbf{w}_{m}^{H}\mathbf{\Lambda}(\theta_{m},l_{m})\mathbf{x}^{(t+1)}|^{2}}.&\\
    	\end{aligned}
    \end{equation}
    It is easy to see that the problem\eqref{MVDR} is equivalent to the well known MVDR problem \cite{G. Cui-2014}, the solution is given by
    \begin{equation}\label{w}
    		\mathbf{w}_{m}^{(t+1)}
    		=\dfrac{(\bm{\Sigma}_{m}(\mathbf{x}^{(t+1)})+\sigma_{n}^{2}\mathbf{I}_{Q})^{-1}\mathbf{\Lambda}(\theta_{m},l_{m})\mathbf{x}^{(t+1)}}{\mathbf{\Upsilon}_{m}^{H}(\bm{\Sigma}_{m}(\mathbf{x}^{(t+1)})+\sigma_{n}^{2}\mathbf{I}_{Q})^{-1}\mathbf{\Upsilon}_{m}},
    \end{equation}
    where $ \mathbf{\Upsilon}_{m}=\mathbf{\Lambda}(\theta_{m},l_{m})\mathbf{x}^{(t+1)} $.\par
    
    Finally, we discuss the update of the auxiliary variables $ \{\mathbf{v}_{p}\}_{p=1}^{M_{\text{jam}}} $ and $ \{\mathbf{g}_{m}\}_{m=1}^{M} $. 
    \subsubsection{Update of $ \mathbf{v}_{p}$}
    Let $ \mathbf{u}_{p}=\mathbf{G}^{H}(\theta_{\text{jam},p})\mathbf{x}-\mathbf{s}_{\text{jam},p}-\mathbf{y}_{p}  $. Then, the optimization problem in\eqref{Eq1:GP-ADMM:sub3} can be rewritten as
    \begin{equation}\label{v_p}
    	\begin{aligned}
    		\min_{\mathbf{v}_{p}}\;& \parallel \mathbf{v}_{p}-\mathbf{u}_{p}\parallel_{2}^{2}&\\
    		\mbox{s.t.}\;& \parallel \mathbf{v}_{p}\parallel_{2}^{2}\leq \varepsilon_{p}. & 
    	\end{aligned}
    \end{equation}
    It can be clearly observed that the solution to\eqref{v_p} is given by 
    \begin{equation}\label{key}
    	\mathbf{v}_{p}=\begin{cases}
    		\mathbf{u}_{p},\;&\parallel \mathbf{u}_{p}\parallel_{2}^{2}\leq \varepsilon_{p},\\
    		\sqrt{\varepsilon_{p}}\frac{\mathbf{u}_{p}}{\parallel \mathbf{u}_{p}\parallel_{2}},\;&\parallel \mathbf{u}_{p}\parallel_{2}^{2}> \varepsilon_{p}.
    	\end{cases}
    \end{equation}
    \subsubsection{Update of $ \mathbf{g}_{m}$}
    For the same reason, we let $ \mathbf{q}_{m}=\mathbf{T}_{m}^{H}\mathbf{x}-\mathbf{t}_{m} $. Then, the optimization problem in\eqref{Eq1:GP-ADMM:sub4} can be formulated by
    \begin{equation}\label{key}
    	\begin{aligned}
    		\min_{\mathbf{g}_{m}}\;& \parallel \mathbf{g}_{m}-\mathbf{q}_{m}\parallel_{2}^{2}&\\
    		\mbox{s.t.}\;& \parallel \mathbf{g}_{m}\parallel_{2}^{2}\geq e_{m}. & 
    	\end{aligned}
    \end{equation}
    The closed-form solution to the above problem is given by
    \begin{equation}\label{key}
    	\mathbf{g}_{m}=\begin{cases}
    		\mathbf{q}_{m},\;&\parallel \mathbf{q}_{m}\parallel_{2}^{2}\geq e_{m},\\
    		\sqrt{e_{m}}\frac{\mathbf{q}_{m}}{\parallel \mathbf{q}_{m}\parallel_{2}},\;&\parallel \mathbf{q}_{m}\parallel_{2}^{2}< e_{m}.
    	\end{cases}
    \end{equation}

    We summarize the steps of the proposed GP-ADMM algorithm in \textbf{Algorithm} \ref{GP_ADMM}. The algorithm terminates when the relative change in the optimization variables falls below a predefined threshold $\iota$, i.e.,
    \begin{equation}\label{key}
    	\parallel \mathbf{x}^{(t+1)}-\mathbf{x}^{(t)}\parallel_{2}+\parallel \mathbf{w}^{(t+1)}-\mathbf{w}^{(t)}\parallel_{2}\leq \iota.
    \end{equation}\par
    
    \begin{algorithm}\label{GP_ADMM}
    	\caption{Proposed GP-ADMM Algorithm for Solving the Problem $ \mathcal{P}_{0,1} $ and $ \mathcal{P}_{0,2}$.}
    	\label{alg1}
    	\LinesNumbered 
    	\KwIn{$ \bm{\Theta}_{\text{jam}} $, $ \{\bar{\theta}_{i}\}_{i=1}^{M+G} $, $ \{\theta_{c,k}\}_{k=1}^{K} $, $ \{\bar{l}_{i}\}_{i=1}^{M+G} $, $ P_{T} $, $ e_{m} $, $ \eta $, $ \varepsilon_{p} $, $ \rho $, $ \lambda $, $ \zeta_{k,p} $, $ \alpha^{(t)} $, $ \iota $.}
    	\KwOut{$ \mathbf{x}^{*},\{\mathbf{w}_{m}^{*}\}_{m=1}^{M} $.}
    	\textbf{Initialize:}$ \; $$ t=0 $, $ \mathbf{x}^{(0)},\{\mathbf{w}_{m}^{(0)}\}_{m=1}^{M},\mathbf{v}_{p}^{(0)},\mathbf{g}_{m}^{(0)} $ and Lagrange multipliers $ \{\mathbf{y}_{p}^{(0)}\} _{p=1}^{M_{\text{jam}}}, \{\mathbf{t}_{m}^{(0)}\} _{m=1}^{M}$\;
    	\Repeat{convergence}{
    		Update of $ \mathbf{x}^{(t+1)} $\\
    		$\mathbf{d}^{(t+1)}=\mathbf{x}^{(t)}-\alpha^{(t)} \nabla_{\mathbf{x}}\mathcal{L}_{\lambda}(\mathbf{x}^{(t)},\mathbf{w}_{m}^{(t)},\mathbf{v}_{p}^{(t)},\mathbf{g}_{m}^{(t)} ,\mathbf{y}_{p}^{(t)},\mathbf{t}_{m}^{(t)})$. \\
    		$ \mathbf{x}^{(t+1)}=\bm{\Pi}_{\bm{\Omega}}\{\mathbf{d}^{(t+1)}\} $.\\
    		Update of $ \{\mathbf{w}_{m}^{(t+1)}\}_{m=1}^{M} $\\
    		\For{$m = 1,2,\dots,M$}{
    			Compute  $ \mathbf{w}_{m}^{(t+1)} $ according to equation\eqref{w}.}
    		Update of $ \{\mathbf{v}_{p}^{(t+1)}\} _{p=1}^{M_{\text{jam}}} $ and $ \{\mathbf{y}_{p}^{(t+1)}\} _{p=1}^{M_{\text{jam}}} $\\
    		\For{$p = 1,2,\dots,M_{\text{jam}}$}{
    			$\mathbf{u}_{p}=\mathbf{G}^{H}(\theta_{\text{jam},p})\mathbf{x}^{(t+1)}-\mathbf{s}_{\text{jam},p}-\mathbf{y}_{p}^{(t)}  $.\\
    			$ \mathbf{v}_{p}^{(t+1)}=\min\{\sqrt{\varepsilon_{p}}/{\parallel \mathbf{u}_{p}\parallel_{2}},1\}\cdot\mathbf{u}_{p}$.\\
    			$ \mathbf{y}_{p}^{(t+1)}=\mathbf{y}_{p}^{(t)}+\mathbf{v}_{p}^{(t+1)}-\mathbf{G}^{H}(\theta_{\text{jam},p})\mathbf{x}^{(t+1)}+\mathbf{s}_{\text{jam},p} $.
    		}
    	    Update of $ \{\mathbf{g}_{m}^{(t+1)}\} _{m=1}^{M} $ and $ \{\mathbf{t}_{m}^{(t+1)}\} _{m=1}^{M} $\\
    	    \For{$m = 1,2,\dots,M$}{
    	    	$\mathbf{q}_{m}=\mathbf{T}_{m}^{H}\mathbf{x}^{(t+1)}-\mathbf{t}_{m}^{(t)}  $.\\
    	    	$ \mathbf{g}_{m}^{(t+1)}=\max\{\sqrt{e_{m}}/{\parallel \mathbf{q}_{m}\parallel_{2}},1\}\cdot\mathbf{q}_{m}$.\\
    	    	$ \mathbf{t}_{m}^{(t+1)}=\mathbf{t}_{m}^{(t)}+\mathbf{g}_{m}^{(t+1)}-\mathbf{T}_{m}^{H}\mathbf{x}^{(t+1)} $.
    	    }
    	}
    	$ \mathbf{x}^{*}=\mathbf{x}^{(t+1)},\{\mathbf{w}_{m}^{*}\}_{m=1}^{M}=\{\mathbf{w}_{m}^{(t+1)}\}_{m=1}^{M}$.
    \end{algorithm}

    \subsection{Computational Complexity Analysis}
    In this section, we briefly analyze the per-iteration computational complexity of the proposed GP-ADMM algorithm.
    It can be seen that the variable $ \mathbf{x} $ is updated in each iteration using the GP step and the BM, where the computational complexity of the gradient of the objective function is mainly related to $ \bm{\Gamma}(\mathbf{w}_{m}^{(t)}) $ and $ \bm{\Psi}(\mathbf{w}_{m}^{(t)}) $. Therefore, the per-iteration complexity of $ \mathbf{x} $ is $ O(N^{2}N_{T}^{2}N_{R}(N+l_{M}-l_{1})+N_{T}N\log_{2}(\sqrt{\eta P_{T}/N_{T}N}/(\epsilon\min\{|d_{i}|,|d_{i}|\neq0\}))) $, where $\epsilon$ denotes the convergence accuracy of the BM. Further, the update of $ \mathbf{w}_{m} $ is mainly computed from the inverse of the matrix $ (\bm{\Sigma}_{m}(\mathbf{x}^{(t+1)})+\sigma_{n}^{2}\mathbf{I}_{N_{R}(N+l_{M}-l_{1})}) $ with computational complexity $ O(N_{R}^{3}(N+l_{M}-l_{1})^{3}) $. By observing, we can see that the auxiliary variables $ \{\mathbf{v}_{p}\} _{p=1}^{M_{\text{jam}}},\{\mathbf{g}_{m}\} _{m=1}^{M} $ and the Lagrange multipliers $ \{\mathbf{y}_{p}\} _{p=1}^{M_{\text{jam}}},\{\mathbf{t}_{m}\} _{m=1}^{M} $ without significant impact on the required computational overhead. Thus, the per-iteration complexity
    of the proposed GP-ADMM algorithm is $ O(N^{2}N_{T}^{2}N_{R}(N+l_{M}-l_{1})+N_{T}N\log_{2}(\sqrt{\eta P_{T}/N_{T}N}/(\epsilon\min\{|d_{i}|,|d_{i}|\neq0\}))+O(N_{R}^{3}(N+l_{M}-l_{1})^{3}) $.

	\subsection{Convergence Analysis}
	We end this section by analyzing the convergence of the proposed algorithm. First, we note that the problems $ \mathcal{P}_{0,1} $ and $ \mathcal{P}_{0,2} $ are solved based on the GP-ADMM algorithm, and the GP framework is mainly derived from the augmented Lagrange functions of $ \mathcal{P}_{0,1} $ and $ \mathcal{P}_{0,2} $. Thus,  we prove the convergence of problems $ \mathcal{P}_{0,1} $ and $ \mathcal{P}_{0,2} $ in the GP framework based on the results in \cite{J. Tranter-2017}. Note that the augmented Lagrange function of\eqref{Lagrange} has a Lipschitz continuous gradient. In the following, we give an important theorem.
	\begin{Theorem}\label{Theorem1}
		The proposed GP-ADMM algorithm converges when the step size chosen for each gradient iteration satisfies $ \alpha^{(t)} <\frac{1}{L^{(t)}}$, where $ L^{(t)} $ is the smallest Lipschitz constant of the function $ \mathcal{L}_{\lambda}(\mathbf{x},\{\mathbf{w}_{m}\}_{m=1}^{M},\{\mathbf{v}_{p}\} _{p=1}^{M_{\text{jam}}},\{\mathbf{g}_{m}\}_{m=1}^{M},\{\mathbf{y}_{p}\} _{p=1}^{M_{\text{jam}}},\{\mathbf{t}_{m}\}_{m=1}^{M}) $ at the $ t $-th iteration.
	\end{Theorem}
	Please see Appendix \ref{appendixA} for the proof. $ \hfill\blacksquare $\par
	
	\begin{Lemma}\label{Lemma1}
		A sequence of iterates generated by the GP-ADMM algorithm, $ \{\mathbf{x},\{\mathbf{w}_{m}\}_{m=1}^{M},\{\mathbf{v}_{p}\}_{p=1}^{M_{\text{jam}}}, \{\mathbf{g}_{m}\}_{m=1}^{M},\{\mathbf{y}_{p}\} _{p=1}^{M_{\text{jam}}} \}$ and $\{\mathbf{t}_{m}\}_{m=1}^{M} $ converges to a KKT point of problem\eqref{Eq1:GP-ADMM}.
	\end{Lemma}
	Please see Appendix \ref{appendixB} for the proof. $ \hfill\blacksquare $\par

	\begin{figure*}[htbp]
		\centering
		\subfigure[]
		{\begin{minipage}{5.9cm}\label{Residual}
				\centering
				\includegraphics[width=1\linewidth]{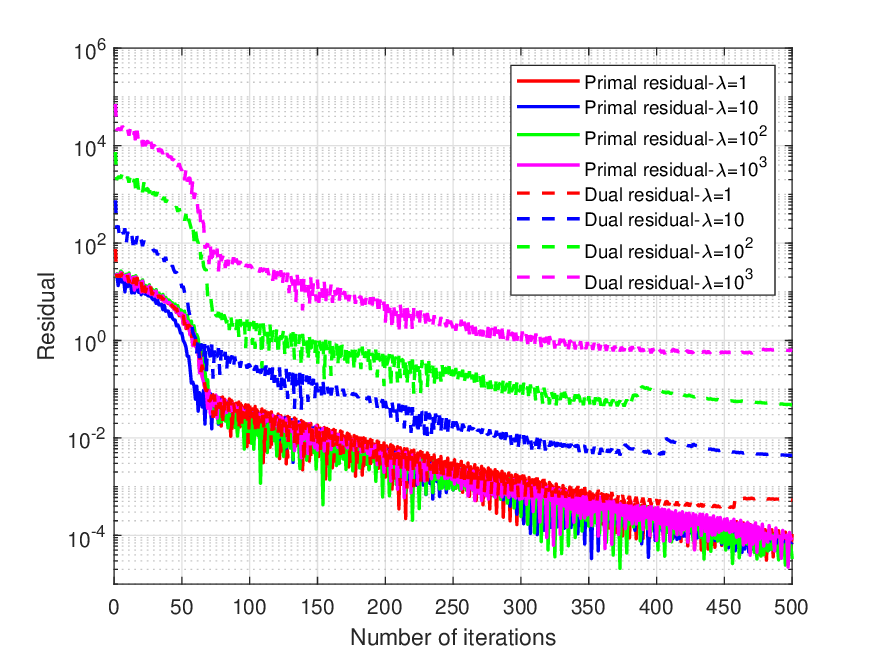}
		\end{minipage}}
		\subfigure[]
		{\begin{minipage}{5.9cm}\label{Obj_lambda_alpha}
				\centering
				\includegraphics[width=1\linewidth]{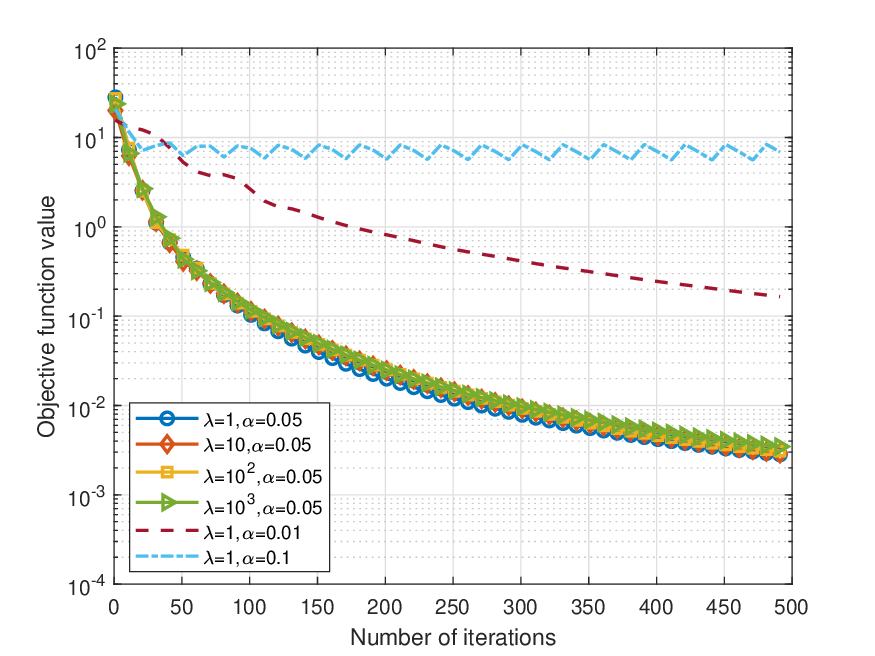}
		\end{minipage}}
		\subfigure[]
		{\begin{minipage}{5.9cm}\label{Obj}
				\centering
				\includegraphics[width=1\linewidth]{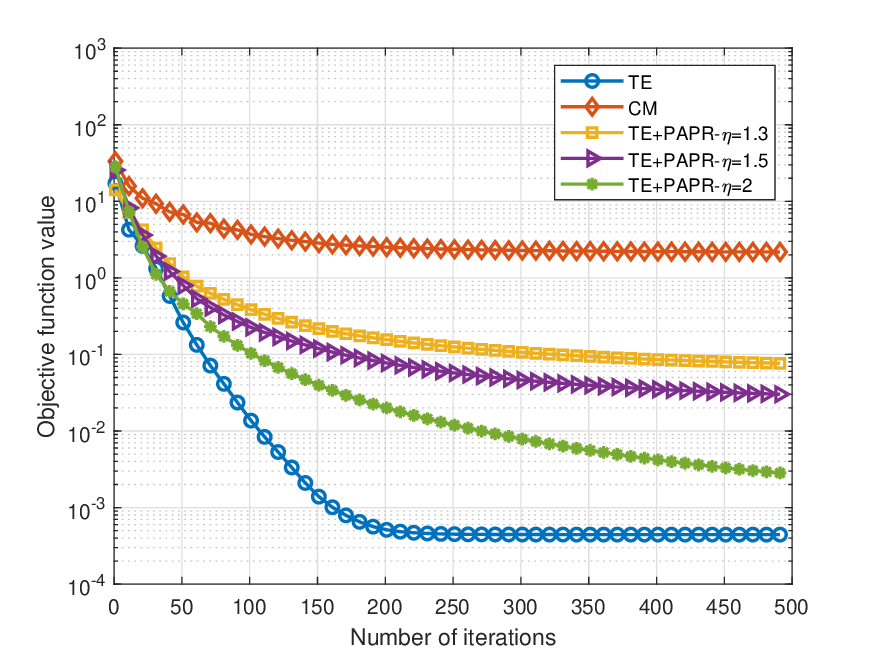}
		\end{minipage}}
		\caption{The convergence curves of the proposed GP-ADMM algorithm. (a) Residuals versus iterations under different $ \lambda $; (b) The objective function values versus iterations under different $ \lambda $  and $ \alpha $; (c) The objective function values versus iterations under different constraints.}	
		\label{Convergence_curves}
	\end{figure*}\par
	\section{Simulation Results}\label{section5}
	In this section,  we demonstrate the performance of the proposed joint transmit waveform and receive filter design method through simulation results. The MIMO MRC system is equipped with $ N_{T}= 16 $ antennas at the transmitter and $ N_{R}= 8 $ antennas at the receiver. The key simulation parameters are set as follows unless specified otherwise. Assume that the MIMO MRC system serves $ K = 5 $ communication users, and it could detects $ M = 2 $ targets, where the user azimuths are $ \theta_{c,1}=-44^{\circ}, \theta_{c,2}=-37^{\circ}, \theta_{c,3}=-23^{\circ}, \theta_{c,4}=41^{\circ} $ and $ \theta_{c,5}=75^{\circ} $ respectively. 
	To reduce the energy leakage from the radar transmitter in target directions, we set $ e_{1}=e_{2}=40 \;\text{W} $. We assume that the directions of ET targets are $ \theta_{\text{jam},1}=-30^{\circ}$ and $\theta_{\text{jam},2}=50^{\circ}$. Furthermore, we report the corresponding target and interference information in Table \ref{target}, where the azimuth errors for the targets and interferences are $ \vartheta_{m}=3^{\circ}, m=1,2 $ and $ \varphi_{g}=3^{\circ}, g=1,2,3 $, respectively.
	The noise variance $ \sigma_{n}^{2}=0 $ dB at the radar receiver and Rician factor $ K_{a}=3 $.\par
	\begin{table}[h]
		\centering
		\caption{Relevant parameters of targets and interference sources }\label{target}
		\setlength{\extrarowheight}{1.5pt}
		\begin{tabular}{|c|c|c|c|}
			\hline
			& \textbf{Mean Azimuth}  &\textbf{Range bin} & \textbf{Power} (dB)\\ \hline
			Target $ 1 $ & $ \bar{\theta}_{1}=-20^{\circ} $ & $ l_{1}=100 $ & $ 10 $ \\
			Target $ 2 $ & $ \bar{\theta}_{2}=20^{\circ} $ & $ l_{2}=109 $ & $ 10 $\\
			Interference $ 1 $ & $ \bar{\phi}_{1}=-50^{\circ} $ & $ \bar{l}_{1}=110 $ & $ 30 $\\
			Interference $ 2 $ & $ \bar{\phi}_{2}=10^{\circ} $ & $ \bar{l}_{2}=115 $ & $ 30 $\\
			Interference $ 3 $ & $ \bar{\phi}_{3}=60^{\circ} $  & $ \bar{l}_{3}=103 $ & $ 30 $\\
			\hline
		\end{tabular}
	\end{table}
	
	For the case of imperfect CSI, assuming that the user angle error $ \Delta\theta_{c} = 2^{\circ} $, $ \omega_{q}=1,q=1,2,\cdots,D $ and $ D=5 $. To ensure the communication performance of the user within the angle uncertainty interval, we sacrifice some radar performance and set $ e_{1}=e_{2}=20\;\text{W} $. Without loss of
	generality, we assume that the perfect communication CSI is used in all simulations unless otherwise stated.

	For the communication purpose, we send quadrature phase shift keying (QPSK) signal to the communication receiver.
	In addition, the system is required to transmit noise-like jamming signals to the ET targets.  In our simulations, the length of the discrete-time transmit waveform is set to $ N=40 $, the upper bound $ \eta=2 $ of PAPR, and the total system transmitted power is $ P_{T} = 40\;\text{W} $. For the ET targets, the power of the desired emitted jamming signals are set to be $  P_{\text{jam},1}=40\;\text{W}$ and $P_{\text{jam},2}=50\;\text{W} $, and the maximum allowed matching errors are $ \varepsilon_{1}=0.02 $ and $ \varepsilon_{2}=0.01 $. In algorithm, to simultaneously ensure the radar, communications and jamming functions, we set a weighting factor $ \rho=0.05 $.
    Moreover, we set the convergence thresholds $ \iota=1\times 10^{-6} $, and the maximum number of iterations is set to $ 5\times 10^{2}$.
	Finally, to avoid the contingency of the results, the performance curves are obtained through $ 10^{3} $ times MonteCarlo simulations.

	\begin{figure}[htbp]
		\centering
		\includegraphics[width=0.9\linewidth]{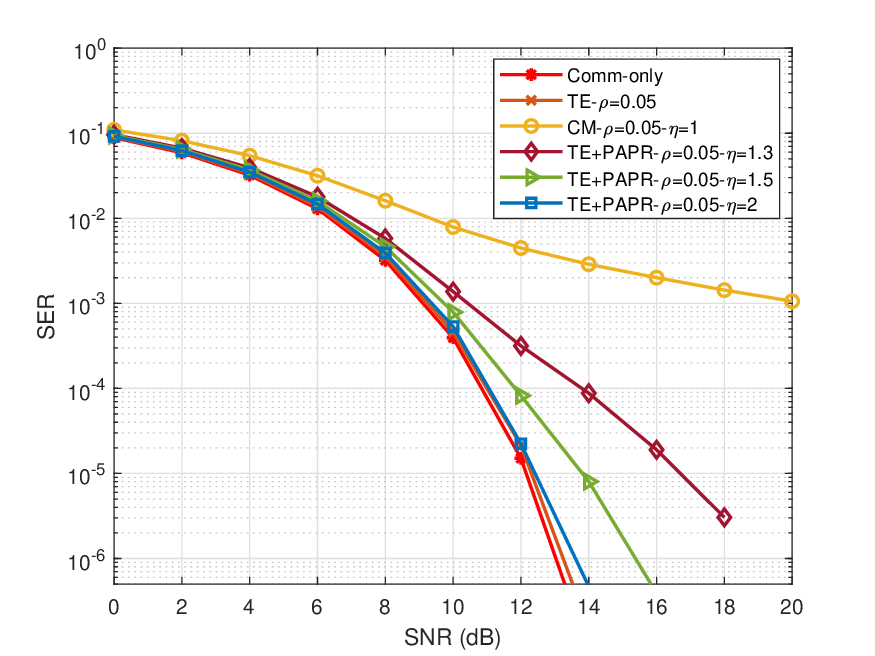}
		\caption{The SER comparison for different constraints.}
		\label{SER}
	\end{figure}
	
	\subsection{Convergence Performance of the GP-ADMM Algorithm}
	The convergence of the proposed algorithm is examined in this section.
	Fig. \ref{Residual} draws the residual curves versus iterations under different penalty parameters $ \lambda $. It can be seen that both the primal and dual residuals show downward trends and eventually converge. Moreover, increasing the penalty parameter $ \lambda $ leaves the primal residuals almost unchanged, while the dual residuals scale up accordingly. In particular, $\lambda=1$ yields more balanced convergence behaviors of the primal and dual residuals, indicating a better tradeoff between primal and dual convergence.
	Fig. \ref{Obj_lambda_alpha} plots the objective function values versus the number of iterations for different values of 
	$ \lambda $ and $ \alpha $. As observed, a relatively large step size ($\alpha=0.1$) leads to divergence of the objective function, whereas a smaller step size ($\alpha=0.01$) results in slower convergence. In contrast, $\alpha=0.05$ provides a favorable balance between convergence speed and stability. Note that the proposed algorithm achieves comparable convergence performance of the objective function under different penalty parameters. Unless otherwise specified, the penalty parameter and step size are set to $\lambda=1$ and $\alpha^{(t)}=0.05$, respectively, in all subsequent simulations to facilitate the convergence of the primal and dual residuals.
	Fig. \ref{Obj} illustrates the curves of the objective function value of the proposed GP-ADMM algorithm with respect to the number of iterations. As it is observed, all of the considered schemes rapidly decrease with iterations until convergence to a stationary point. It is worth highlighting that the TE constraint involved converge to the stationary points below the CM constraint (i.e.,
	waveforms with PAPR $ \eta = 1 $). This is because the CM constraint is stronger than the TE constraint, leading to the feasible set of solutions is smaller. Moreover, in the case of the TE+PAPR constraint, it can be seen
	that the objective function value decreases with the increase of the PAPR of the waveform. The results indicate that the proposed method can effectively optimize the transmit waveform and the radar receive filter in the case of the different constraints.

	\subsection{Communication and Radar Performance} 
	This section evaluates the performance of communication and radar of the MIMO MRC system. We start with the evaluation of the communication symbol error rate (SER) of system. The transmit signal-to-noise ratio (SNR) is defined as $\mathbb{E}\{\parallel \mathbf{x}\parallel_{2}^{2}\}/\sigma_{z}^{2}$. In Fig. \ref{SER}, the SER achieved by different constraints versus the transmit SNR is depicted.
	It can be seen that the performance of the proposed method under TE constraint is almost identical to that of the communication only (Comm-only). Note that the SER is worse for approaches CM constraint compared to the TE and TE+PAPR constraints. This is the case, since the strict structure imposed on the transmitted waveform by the CM constraint.
	Furthermore, the SER is decreased with the increase of the PAPR upper bounds $ \eta $ of waveforms. This is due to the fact that the increase in $ \eta $ raises the feasible set of solutions to the problem $ \mathcal{P}_{0,2} $. This is also consistent with the convergence results discussed in Fig. \ref{Obj}.

	\begin{figure*}[htbp]
		\centering
		\subfigure[]
		{\begin{minipage}{5.9cm}\label{ASR_rho}
				\centering
				\includegraphics[width=1\linewidth]{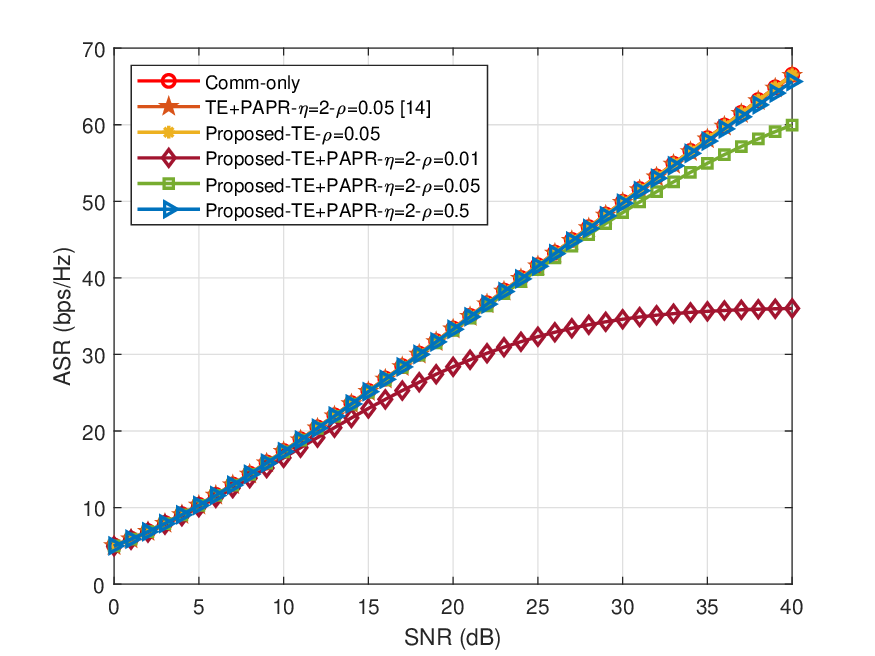}
		\end{minipage}}
		\subfigure[]
		{\begin{minipage}{5.9cm}\label{ASR_PAPR}
				\centering
					\includegraphics[width=1\linewidth]{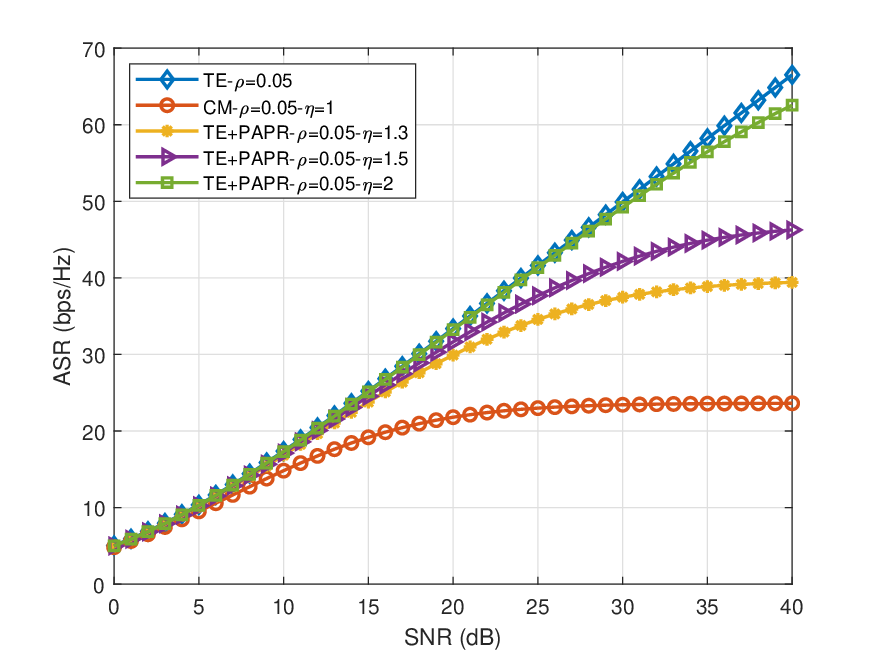}
		\end{minipage}}
	    \subfigure[]
	    {\begin{minipage}{5.9cm}\label{ASR_ImCSI}
	    		\centering
	    		\includegraphics[width=1\linewidth]{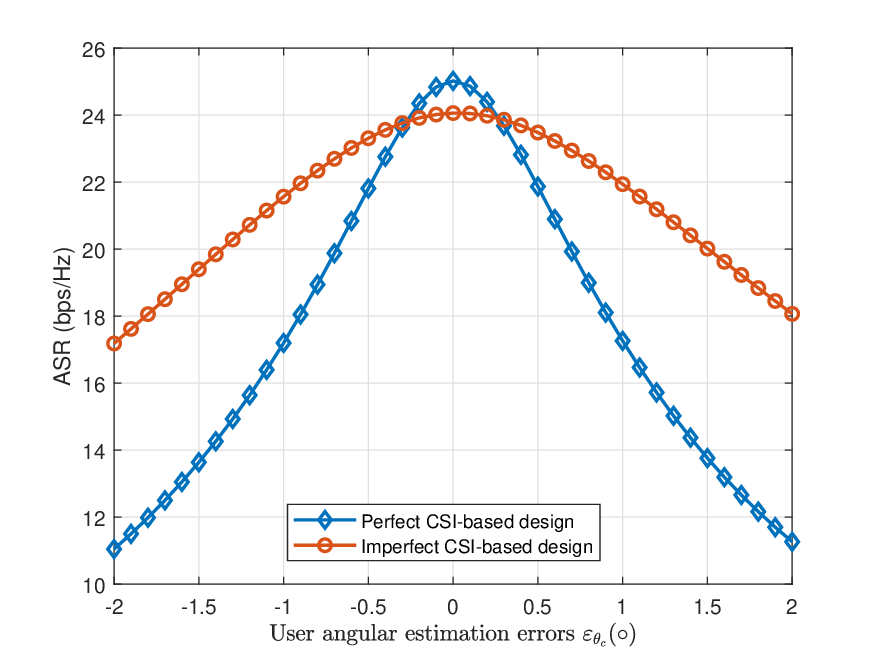}
	    \end{minipage}}
		\caption{The communication ASR for MIMO MRC system. (a) TE and TE+PAPR constraints under different weighting factors $ \rho $; (b) The different PAPR upper bounds $ \eta $ of waveform; (c) Estimation errors $\varepsilon_{\theta_{c}}$ from different user angular (TE+PAPR-$ \rho=0.05,\eta=2 $, communication SNR $=15$ dB).}		
	\end{figure*}\par

    \begin{figure*}[htbp]
    	\centering
    	\subfigure[]
    	{\begin{minipage}{5.9cm}\label{Beam_rho}
    			\centering
    			\includegraphics[width=1\linewidth]{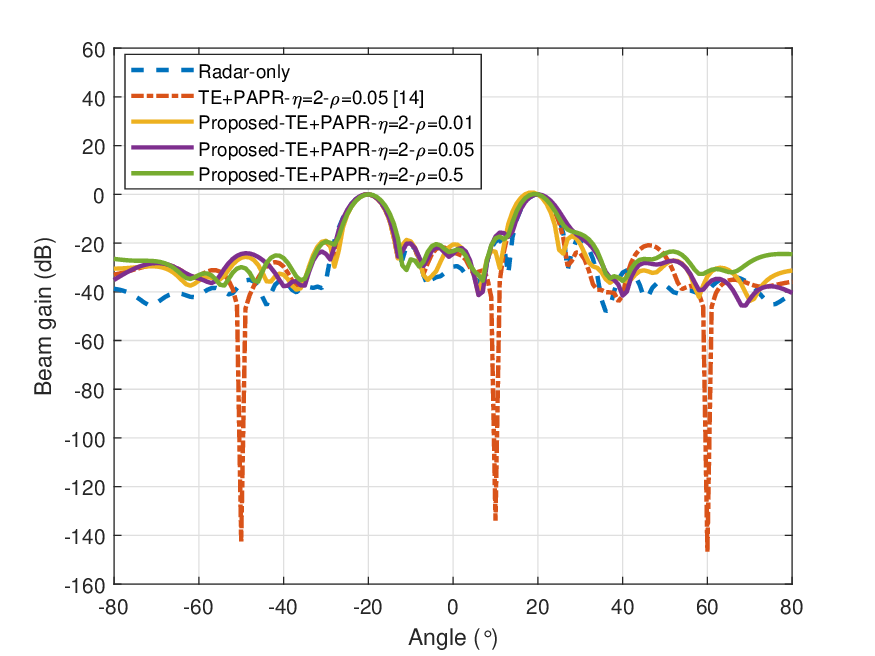}
    	\end{minipage}}
        \subfigure[]
    	{\begin{minipage}{5.9cm}\label{Beam_M}
    			\centering
    			\includegraphics[width=1\linewidth]{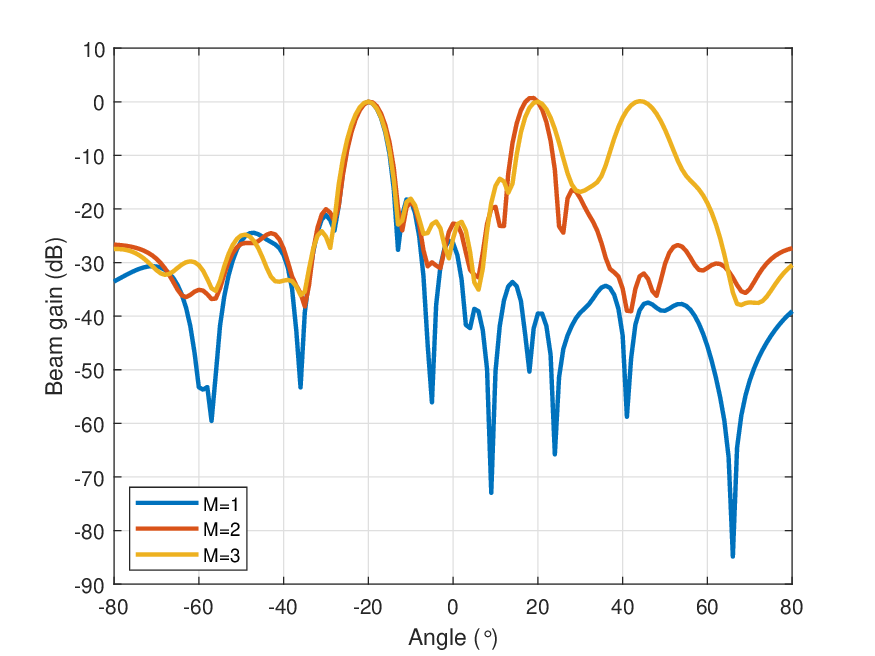}
    	\end{minipage}}
        \subfigure[]
    	{\begin{minipage}{5.9cm}\label{Beam_ImCSI}
    			\centering
    			\includegraphics[width=1\linewidth]{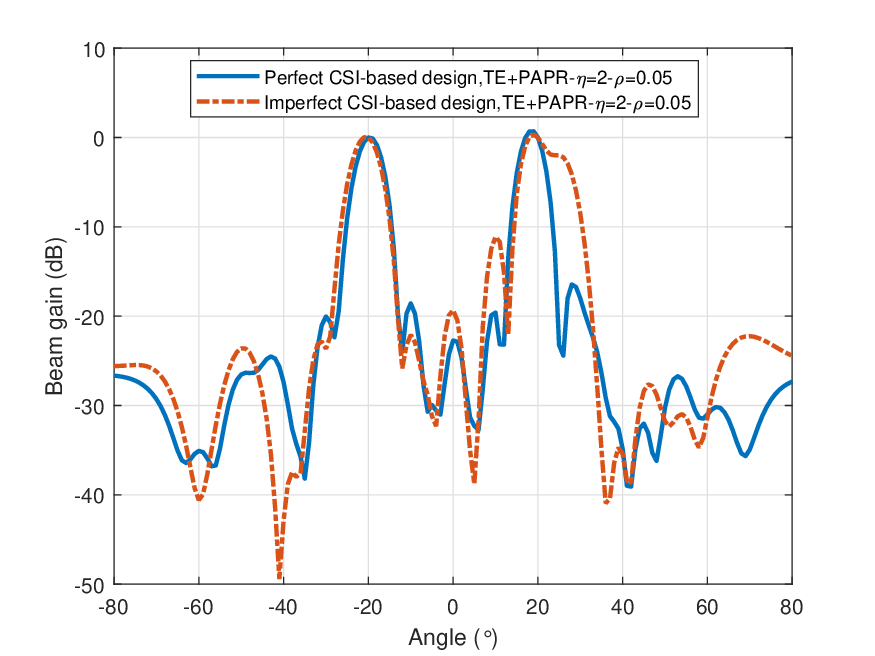}
    	\end{minipage}}
    	\caption{The beampatterns for MIMO MRC system. (a) The different weighting factors $ \rho $; (b) The different the number of target; (c) The imperfect CSI.}	
    	\label{Beam}
    \end{figure*}\par
    
    \begin{figure*}[htbp]
    	\centering
    	\subfigure[]
    	{\begin{minipage}{8cm}\label{CAF1}
    			\centering
    			\includegraphics[width=1\linewidth]{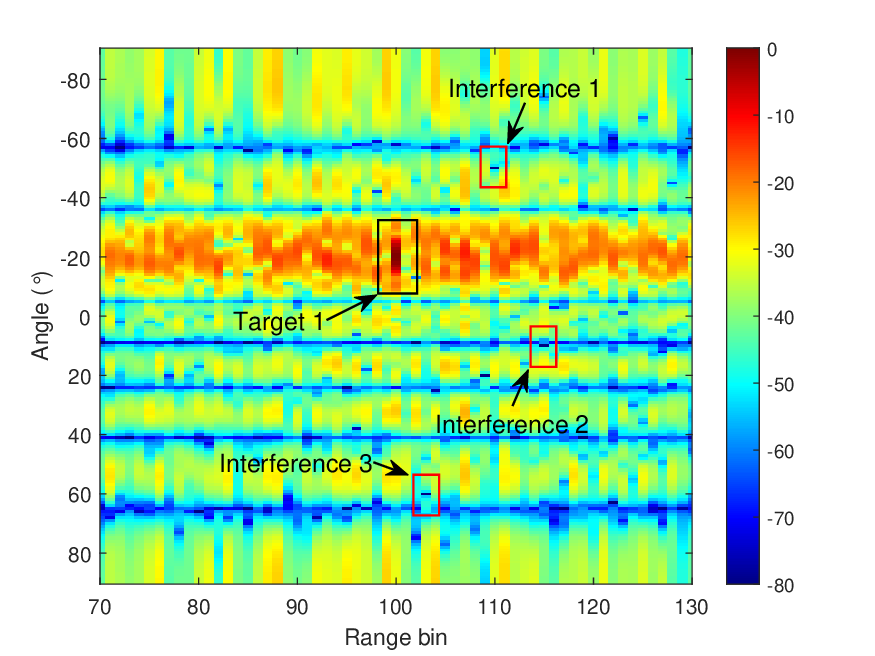}
    	\end{minipage}}
    	\subfigure[]
    	{\begin{minipage}{8cm}\label{CAF2}
    			\centering
    			\includegraphics[width=1\linewidth]{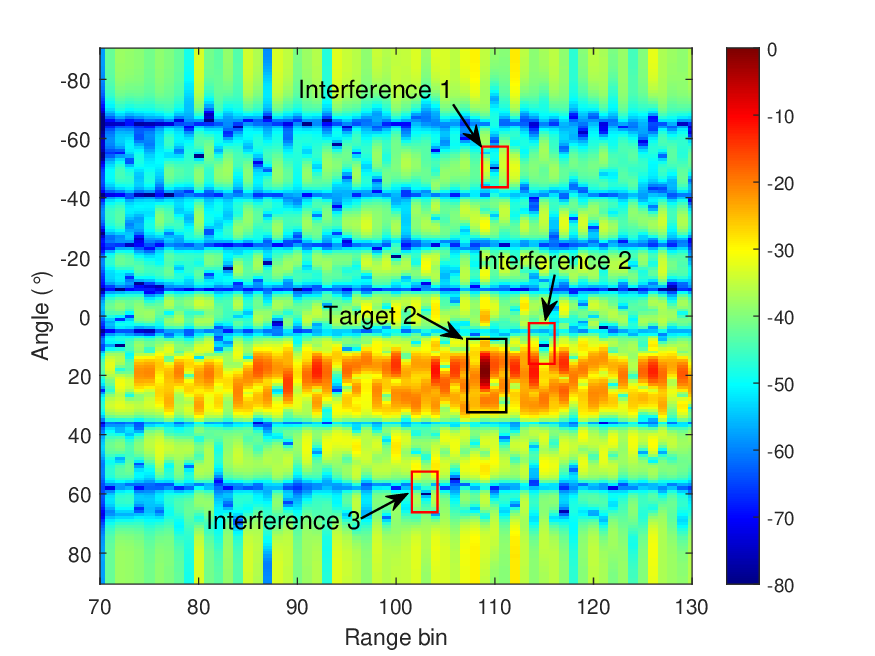}
    	\end{minipage}}
    	\caption{Contour maps of range-angle planes of CAF. (a) $ \mathbf{x} $ and $ \mathbf{w}_{1} $; (b) $ \mathbf{x} $ and $ \mathbf{w}_{2} $.}
    \end{figure*}\par

	In the following simulation experiment, we analyze the communication ASR of the system.
	Figs. \ref{ASR_rho} and \ref{ASR_PAPR} show the ASR versus the transmit SNR under different weighting factors $\rho$ and PAPR upper bounds $\eta$, respectively, while Fig. \ref{ASR_ImCSI} shows the communication ASR under different user angle errors $\varepsilon_{\theta_{c}}$.
	In Fig. \ref{ASR_rho}, we compare the proposed MRC design method with the Comm-only and the ISAC design method proposed in \cite{C. G. Tsinos-2021}. When the SNR is below $ 20 $ dB, the communication ASR of the proposed method closely matches that of the benchmark method in \cite{C. G. Tsinos-2021}, the ASR decreases only at high SNRs (e.g., above $ 25 $ dB). The results demonstrate that the proposed method achieves a better overall performance balance at the cost of only a slight communication performance loss, while significantly enhancing the integrated multifunctional performance of the system.
	Moreover, it can be observed that the scenario with the TE constraint achieves a higher communication ASR than that with the TE+PAPR constraint. We can see that a decrease in the  $ \rho $ value causes a decrease in the communication ASR for the TE+PAPR constraint. This is because a low $ \rho $, will reduce the weight of the system on the communication part of the optimization. It is worth highlighting that when we set $ \rho= 0.05$, the communication performance under TE constraint is close to that Comm-only. Fig. \ref{ASR_PAPR} depicts the ASR under different PAPR upper bounds $ \eta $. For comparison, we consider the case of the TE and CM constraints. 
	It can be seen that the CM constraint has severe impact on the communication ASR of the MIMO MRC system. In addition, the communication ASR decreases significantly with the reduce of the PAPR of the transmit waveform. Therefore, if an ideal PAPR is sought, significant degradation is observed on the ASR. Of course, one may actual requirements by appropriately tuning the $ \eta $. 

	Fig. \ref{ASR_ImCSI} illustrates the impact of user angular estimation errors $ \varepsilon_{\theta_{c}} $ on the communication ASR under perfect and imperfect CSI.
	It can be observed that as the estimation error $ \varepsilon_{\theta_{c}} $ increases, the ASR under perfect CSI degrades significantly, whereas that under imperfect CSI remains relatively stable.	
	Although perfect CSI achieves a higher ASR when the estimation error is small, the imperfect CSI-based design exhibits better robustness as the error increases, demonstrating its effectiveness in mitigating the impact of angular uncertainties.
	\par

    We now move to the evaluation of the radar performance of the  MIMO MRC system. In Fig. \ref{Beam}, the beampatterns of designed waveform for proposed multi-target MIMO MRC scenario are shown. The different weighting factors $ \rho $ are simulated in Fig. \ref{Beam_rho}, it is evident that as $ \rho $ decreases, the beam gain of the main lobes towards the target directions get higher. And the gain of side lobes get lower, which
    makes the power focus on the target directions and enhances the detection ability. However, the side lobes are not reduced blindly, high gain phenomenon still exists in some non-target directions, for the reason that the MIMO MRC system needs the side lobes to communication and jamming. Furthermore, the beampattern generated by the method in \cite{C. G. Tsinos-2021} is compared. The beam gain of the proposed method towards the target directions is comparable to that of \cite{C. G. Tsinos-2021}. Nevertheless, the underlying design philosophies differ significantly. While the method in \cite{C. G. Tsinos-2021} concentrates power toward the target directions and simultaneously forms nulls in the spatial directions of the interference to suppress it, the proposed method is designed to focus on both the spatial directions and range bins of the interference. Finally, it is worth noting that the system can achieve the optimal beam gain at the target when operating in the radar-only mode. The beampatterns for different target numbers are shown in Fig. \ref{Beam_M}. It can be seen that as the number of target increases, the beam gain in target directions decreases, which is due to the fact that the system power is allocated more at the target directions.
    Fig. \ref{Beam_ImCSI} demonstrates the radar beampatterns for the perfect and imperfect CSI cases. Compared to the perfect CSI, the beam gain under imperfect CSI is significantly lower. The reason is that more resources are allocated to optimize the communication performance within the user angular uncertainty region, resulting in a decrease in radar performance.\par

    We also study the joint range and azimuth behavior by using the cross ambiguity function (CAF) \cite{X. Yu-2020} of $ \mathbf{x} $ and $ \mathbf{w}_{m} $ for the $ m $-th target of interest. We assume that a target is located at range bin $ l_{m} $ and angle $ \theta_{m} $. Then, the CAF is defined as
    \begin{equation}\label{key}
    	\mathcal{A}(\mathbf{x},\mathbf{w}_{m},l,\theta)=|\mathbf{w}_{m}^{H}(\mathbf{J}_{l-l_{1}}^{T}\otimes \mathbf{A}(\theta))\mathbf{x}|^{2},
    \end{equation}
    where $ m=1 $ and $ 2 $, and $ l=l_{1}-N+1,\cdots,l_{1}+N-1$.\par
    
    Fig. \ref{CAF1} and Fig. \ref{CAF2} depict the contour maps of range-angle plane of  $ \mathcal{A}(\mathbf{x},\mathbf{w}_{1},l,\theta) $ and $ \mathcal{A}(\mathbf{x},\mathbf{w}_{2},l,\theta) $, respectively. We can see that a prominent peak is observed within the range bin $ l_{m} $ and angle range $ [\bar{\theta}_{m}-\vartheta_{m}/2,\bar{\theta}_{m}+\vartheta_{m}/2] $ which is corresponding to the target, while very low values are achieved at the signal-dependent interference sources (marked by red rectangles). These results demonstrate that the proposed method yields a reasonable CAF for target detection under signal-dependent interference.

    \begin{figure}[htbp]
    	\centering
    	\includegraphics[width=0.9\linewidth]{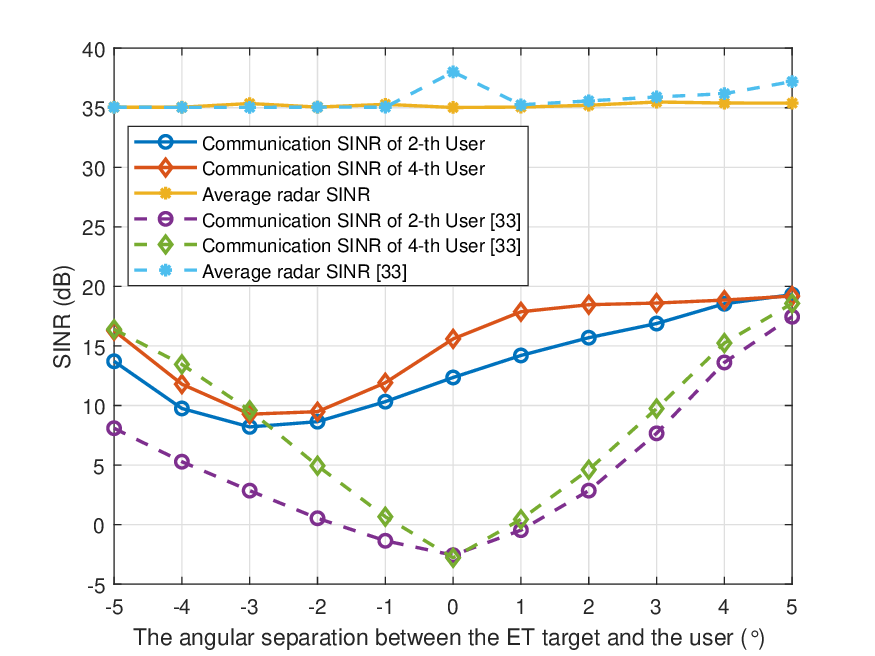}
    	\caption{Radar and communication SINR versus angular separation between the ET target and the user (communication SNR $ = 20 $ dB, $ \rho=0.5 $).}
    	\label{SINR_rad_com_ET}
    \end{figure}
    To analyze the impact of the ET targets' locations relative to users on the communication and radar performance, Fig. \ref{SINR_rad_com_ET} shows the communication and average radar SINR for two ET targets located near the $ 2 $-th and $ 4 $-th users, respectively, at different angular separations $ \Delta\theta =  \theta_{\mathrm{jam},p}-\theta_{c,k},p=1,k=2$ and $p=2,k=4  $ (provided that the jamming constraints are satisfied). When the two directions are well separated, both the proposed method and \cite{B. Tang-2022} achieve satisfactory performance. However, as $ \Delta\theta $ approaches $ 0^\circ$, the communication SINR degrades significantly for method proposed in \cite{B. Tang-2022}. 
    This is because closely spaced ET targets and users cause the jamming beam and communication channel to become spatially correlated, making it difficult to satisfy the communication constraint and the jamming constraint simultaneously. Compared with \cite{B. Tang-2022}, the proposed method suffers less performance loss and maintains a more stable radar SINR across the entire angular range, demonstrating its better ability to handle this spatial conflict.

    \begin{figure}[htbp]
    	\centering
    	\subfigure[]
    	{\begin{minipage}{8cm}\label{Average_SINR_com}
    			\centering
    			\includegraphics[width=1\linewidth]{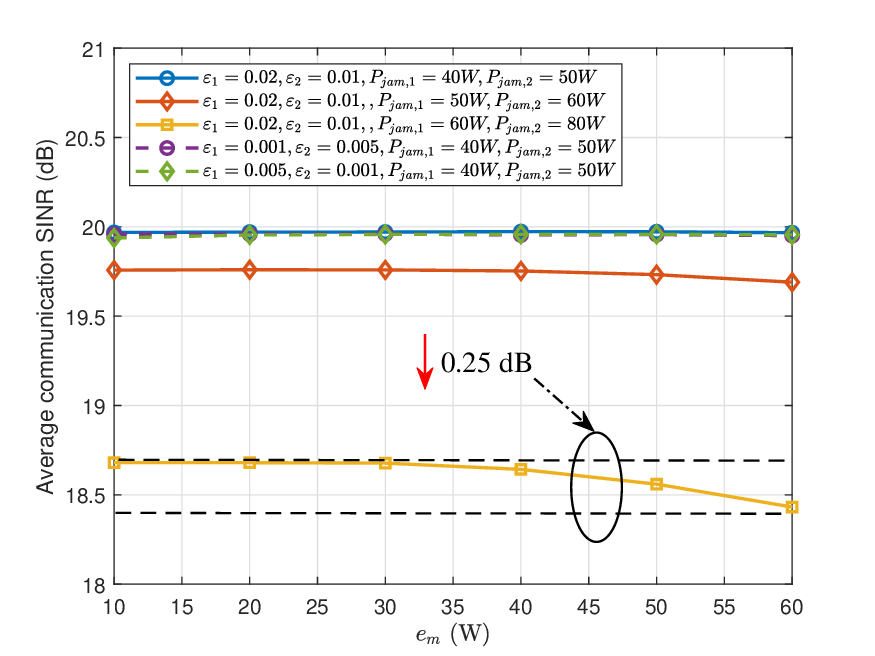}
    	\end{minipage}}
    	\subfigure[]
    	{\begin{minipage}{8cm}\label{Average_SINR_rad}
    			\centering
    			\includegraphics[width=1\linewidth]{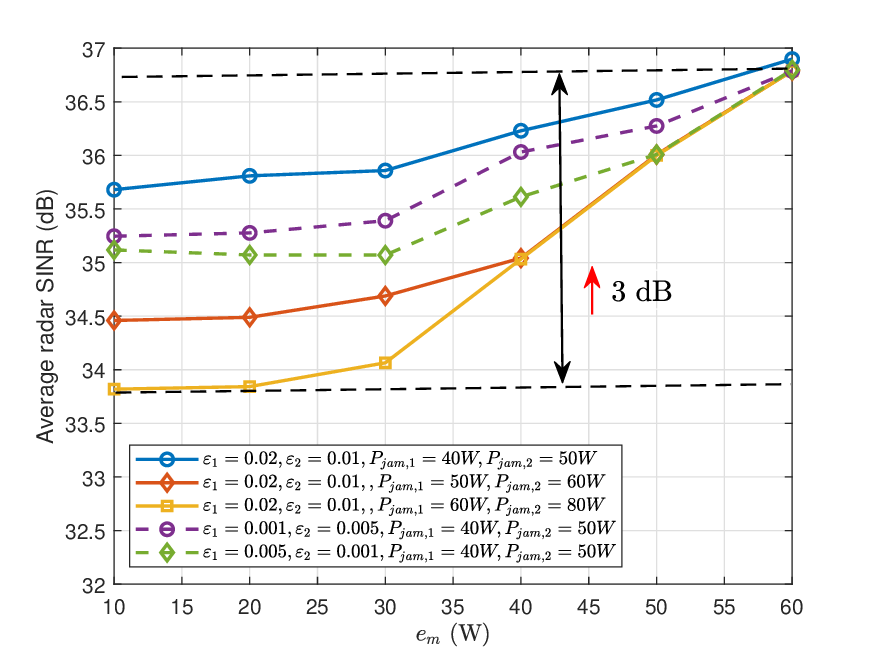}
    	\end{minipage}}
    	\caption{Impact of jamming power, jamming constraint threshold and  transmit energy toward target on the performance of communication and radar ($ \rho=0.5 $). (a) The average communication SINRs (communication SNR $ = 20 $ dB); (b) The average radar SINRs.}			
    \end{figure}\par
	To evaluate the joint trade-off among target detection, downlink communication, and electronic jamming, Figs. \ref{Average_SINR_com} and \ref{Average_SINR_rad} illustrate average communication and radar SINRs, respectively, under different jamming power requirements $ P_{\text{jam},p} $, jamming matching error bounds $ \varepsilon_{p}$ and target-directed transmit energy $ e_{m}$, where $p=1,2$ and $m=1,2,\ldots,M$. 
	As shown in Fig. \ref{Average_SINR_com}, increasing $ e_{m}$ generally degrades the communication SINR, with a more pronounced effect under demanding jamming requirements. For instance, when $P_{\text{jam},1}=60\;\text{W}, P_{\text{jam},2}=80\;\text{W}$ and  $ \varepsilon_{1}=0.02,\varepsilon_{2}=0.01 $, increasing $ e_{m} $ from $ 10\;\text{W} $ to $ 60\;\text{W} $ results in a noticeable communication SINR degradation of approximately $ 0.25 $ dB. This degradation stems from the limited transmit energy budget, under which allocating more energy to target direction reduces the resources available for communication and jamming. In contrast, under relatively stringent matching error constraints ($\varepsilon_{1}=0.005$ and $\varepsilon_{2}=0.0001$) with lower jamming power requirements, the communication SINR exhibits less sensitivity to $ e_{m} $ and maintains a higher performance level.
	Fig. \ref{Average_SINR_rad} further depicts the radar performance. As expected, increasing $ e_{m}$ significantly enhances the average radar SINR, as more energy is forced to be concentrated in the target directions to minimize energy leakage. Notably, for the most demanding jamming case, the radar SINR achieves a substantial gain of approximately $ 3 $ dB as $ e_{m} $ increases, despite starting from the lowest baseline. Specifically, when the jamming task requires high power and precise waveform matching, it severely depletes the system resources, thereby impairing both communication and radar baselines. Forcing a larger $ e_{m} $ in this scenario effectively reclaims energy from the jamming function back to the radar function, restoring the detection capability at the cost of communication quality. These results validate the effectiveness of the proposed joint waveform design in navigating complex multi-functional trade-offs. Nevertheless, under a limited transmit energy budget, excessively stringent target energy requirements may lead to infeasibility of the optimization problem. Accordingly, the value of 
	$ e_{m}$ should be selected by striking a reasonable tradeoff in practice.

	\subsection{Jamming Performance}
    To evaluate the ECM capability of the MIMO MRC system, this section analyzes the jamming performance of the jamming signals synthesized by the system. \par
	
\begin{figure}[htbp]
		\centering
		\subfigure[]
		{\begin{minipage}{4.3cm}\label{Desired_Synthesized_jamming}
				\centering
				\includegraphics[width=1\linewidth]{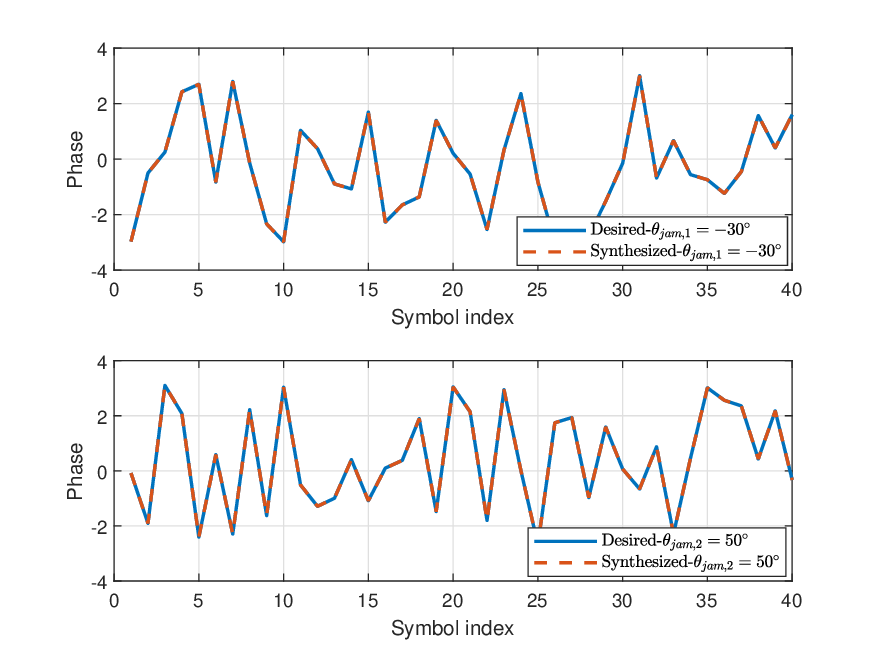}
		\end{minipage}}
		\subfigure[]
		{\begin{minipage}{4.3cm}\label{Pd_real_imag}
				\centering
				\includegraphics[width=1\linewidth]{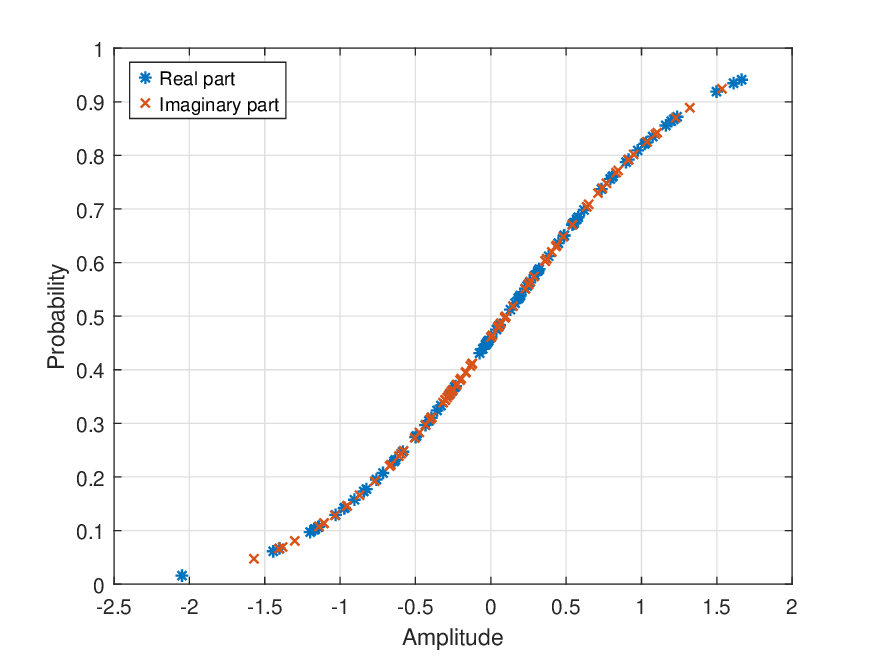}
		\end{minipage}}
		\caption{The synthesized jamming signal. (a) Phase comparison between the desired jamming signal and the synthesized jamming signal; (b) The normal probability of the synthesized jamming signal.}			
	\end{figure}

	Fig. \ref{Desired_Synthesized_jamming} plots the desired transmit noise-like jamming signals and the synthesized jamming signals of system at $ \theta_{\text{jam},1}=-30^{\circ}$  and $\theta_{\text{jam},2}=50^{\circ}$.
	We can see that the phase of the synthesized jamming signals approximates the desired. 
	The normal probability of the real and imaginary parts of the synthesized jamming signal are plotted in Fig. \ref{Pd_real_imag}. The results indicate that the synthesized signals follow a noise-like property as desired.  In summary, the MIMO MRC system is capable of radar and communications while transmitting noise-like signals to ET targets.
	
	\par
    


	\begin{figure}[htbp]
		\centering
		\subfigure[]
		{\begin{minipage}{8cm}\label{SER_jamming}
				\centering
				\includegraphics[width=1\linewidth]{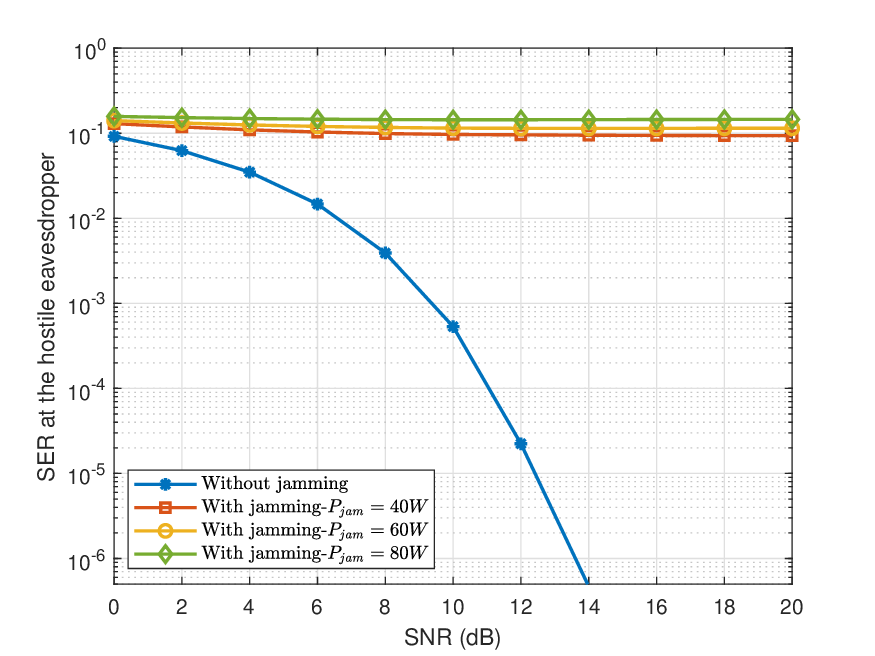}
		\end{minipage}}
		\subfigure[]
		{\begin{minipage}{8cm}\label{Pd_jamming}
				\centering
				\includegraphics[width=1\linewidth]{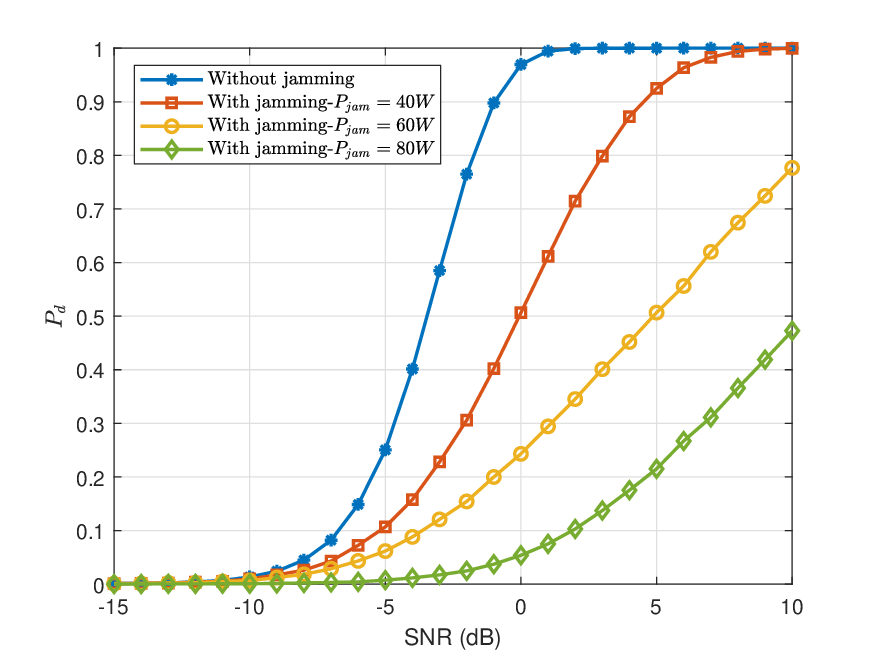}
		\end{minipage}}
		\caption{The ECM capability of the synthesized jamming signal. (a) The communication SER; (b) The probability of detection ($ P_{fa}=10e^{-6} $).}				
	\end{figure}\par
     To further evaluate the ECM capability of the synthesized jamming signals, we consider two hostile receivers, namely, a communication eavesdropper and a hostile radar. For the communication eavesdropper, the receiver is assumed to employ a maximum likelihood detector after the Gaussian channel under perfect synchronization.
     For the hostile radar, a linearly frequency modulated (LFM) waveform with a pulse duration of $ 10\;\mu s $ and a bandwidth of $ 2\;\text{MHz} $ is considered. The received signal consists of the target echo, the jamming signal transmitted by the MIMO MRC system, and additive noise. The hostile radar employs a matched-filter-based detector with coherent integration. The detection threshold is determined to satisfy a false alarm probability of $ (P_{fa}=10e^{-6}) $.
     The without jamming case considers only the target echo and noise at the hostile receiver, whereas the with jamming case additionally incorporates the jamming signal generated by the proposed MIMO MRC system and the jamming-to-signal-ratio is $ 0 $ dB. Different values of $ P_{\text{jam}}$ are considered to investigate the impact of the jamming power on the hostile receivers.

	 Fig. \ref{SER_jamming} shows the SER of the hostile communication receiver.  It can be observed that when hostile eavesdropper is jammed by jamming signals from the MIMO MRC system, its communication receivers exhibit a sharp degradation in SER.
	 In particular, increasing the jamming power further elevates the SER of eavesdropper, effectively thwarting its ability to intercept communication data. This confirms that the proposed method successfully safeguards against eavesdropping.
	 Fig. \ref{Pd_jamming} depicts the target detection probability of the hostile radar versus SNR, comparing scenarios before and after jamming. The results demonstrate that the jamming signal released by the system significantly degrades the target detection probability of the hostile radar. Moreover, increasing the jamming power further suppresses the detection probability of the hostile radar, enhancing the jamming effectiveness.

	\section{Conclusion}\label{section6}
	This paper investigated a MIMO MRC system for simultaneously enhancing the beam gain for targets detection, jam ET targets, and maintain reliable communications. We demonstrated that the ECM capability of such a system is fundamentally governed by the energy of the transmitted jamming signal.
	To address these objectives, we proposed a joint optimization framework for transmit waveform and receive filter design, and considered the perfect and imperfect communication CSI, respectively. 
	Two practical constraints were considered: TE constraint and combined TE+PAPR constraint. For each case, we formulated a multi-objective non-convex trade-off optimization problem to jointly maximize radar receive SINR and minimize communication MUI energy under jamming functionality and directional target energy constraints. 
	To solve these challenging problems, we developed an efficient GP-ADMM algorithm, deriving closed-form solutions for all subproblems. The convergence of the proposed algorithm to a KKT point was rigorously analyzed and theoretically proved. Simulation results validated that the proposed method can be used to achieve satisfactory performance of communications and radar while effectively preventing eavesdropping of communications data by the ET targets and degrading the detection probability of hostile radar.

	\appendices
		\section{proof of Theorem \ref{Theorem1}}\label{appendixA}
		For simplicity, we omit the superscript of variables $ \{\mathbf{w}_{m}\}_{m=1}^{M},\{\mathbf{v}_{p}\} _{p=1}^{M_{\text{jam}}},\{\mathbf{g}_{m}\}_{m=1}^{M},\{\mathbf{y}_{p}\} _{p=1}^{M_{\text{jam}}}$ and $\{\mathbf{t}_{m}\}_{m=1}^{M}$ in the following proof.\par
		
		$ Proof $: We first analyze the convergence of the Lagrange multipliers $ \mathbf{y}_{p} $ and $ \mathbf{t}_{m}$ in the ADMM framework. Since the objective function $ f(\mathbf{x},\{\mathbf{w}_{m}\}_{m=1}^{M}) $ is bounded, we have the following conclusion holds when $ t\to\infty $
		\begin{equation}\label{key}
			\lim_{t\to\infty}\parallel\mathbf{v}_{p}^{(t+1)}-\mathbf{G}^{H}(\theta_{\text{jam},p})\mathbf{x}^{(t+1)}+\mathbf{s}_{\text{jam},p} \parallel_{2}^{2}=0,
		\end{equation}
	    \begin{equation}\label{key}
	    	\lim_{t\to\infty}\parallel\mathbf{g}_{m}^{(t+1)}-\mathbf{T}_{m}^{H}\mathbf{x}^{(t+1)}\parallel_{2}^{2}=0,
	    \end{equation}
        Further, we have
		\begin{equation}\label{proof_v_p_lim}
			\lim_{t\to\infty}(\mathbf{v}_{p}^{(t+1)}-\mathbf{G}^{H}(\theta_{\text{jam},p})\mathbf{x}^{(t+1)}+\mathbf{s}_{\text{jam},p} )=\mathbf{0}_{N\times 1}.
		\end{equation}
	    \begin{equation}\label{key}
	    	\lim_{t\to\infty}(\mathbf{g}_{m}^{(t+1)}-\mathbf{T}_{m}^{H}\mathbf{x}^{(t+1)})=\mathbf{0}_{N\times 1}.
	    \end{equation}
		Hence, by definition of the Lagrange multipliers $ \mathbf{y}_{p} $ and $ \mathbf{t}_{m}$, we have
		\begin{equation}\label{proof_y_p}
			\mathbf{y}_{p}^{(t+1)}-\mathbf{y}_{p}^{(t)}=\mathbf{v}_{p}^{(t+1)}-\mathbf{G}^{H}(\theta_{\text{jam},p})\mathbf{x}^{(t+1)}+\mathbf{s}_{\text{jam},p}.
		\end{equation}
	    \begin{equation}\label{key}
	    	\mathbf{t}_{m}^{(t+1)}-\mathbf{t}_{m}^{(t)}=\mathbf{g}_{m}^{(t+1)}-\mathbf{T}_{m}^{H}\mathbf{x}^{(t+1)}.
	    \end{equation}
		Combining\eqref{proof_v_p_lim} and\eqref{proof_y_p}, we can obtain
		\begin{equation}\label{key}
			\lim_{t\to\infty}(\mathbf{y}_{p}^{(t+1)}-\mathbf{y}_{p}^{(t)} )=\mathbf{0}_{N\times 1}.
		\end{equation}
		Thus, the Lagrange multiplier $ \mathbf{y}_{p} $ is convergent. To this end, we suppose that  $ \mathbf{y}_{p} $ converges to $ \mathbf{y}_{p}^{*} $, i.e.,
		\begin{equation}\label{key}
			\lim_{t\to\infty}\mathbf{y}_{p}^{(t+1)}=\lim_{t\to\infty}\mathbf{y}_{p}^{(t)}=\mathbf{y}_{p}^{*}.
		\end{equation}
	    Similarly, we can prove that the Lagrange multiplier $ \mathbf{t}_{m} $ is convergent, and suppose that  $ \mathbf{t}_{m} $ converges to $ \mathbf{t}_{m}^{*} $, i.e., $ \lim\limits_{t\to\infty}\mathbf{t}_{m}^{(t+1)}=\lim\limits_{t\to\infty}\mathbf{t}_{m}^{(t)}=\mathbf{t}_{m}^{*} $.
	    
		Then, the updates of $ \mathbf{x},\mathbf{w}_{m},\mathbf{v}_{p}$ and $ \mathbf{g}_{m} $ can be computed through
		handling the following subproblems
		\begin{subequations}\label{Eq2:GP-ADMM}
			\begin{align}
				\mathbf{x}^{(t+1)}&=\arg \min_{\mathbf{x}\in \bm{\Omega}} \mathcal{L}_{\lambda}(\mathbf{x},\mathbf{w}_{m}^{(t)},\mathbf{v}_{p}^{(t)},\mathbf{g}_{m}^{(t)},\mathbf{y}_{p}^{*},\mathbf{t}_{m}^{*} ),\label{Eq2:GP-ADMM:sub1}\\
				\mathbf{w}_{m}^{(t+1)}&=\arg \min_{\mathbf{w}_{m}} \mathcal{L}_{\lambda}(\mathbf{x}^{(t+1)},\mathbf{w}_{m},\mathbf{v}_{p}^{(t)},\mathbf{g}_{m}^{(t)},\mathbf{y}_{p}^{*},\mathbf{t}_{m}^{*}),\label{Eq2:GP-ADMM:sub2}\\
				\mathbf{v}_{p}^{(t+1)}&=\arg\min_{\mathbf{v}_{p}} \mathcal{L}_{\lambda}(\mathbf{x}^{(t+1)},\mathbf{w}_{m}^{(t+1)},\mathbf{v}_{p},\mathbf{g}_{m}^{(t)},\mathbf{y}_{p}^{*},\mathbf{t}_{m}^{*} ),\label{Eq2:GP-ADMM:sub3}\\
				\mathbf{g}_{m}^{(t+1)}&=\arg\min_{\mathbf{g}_{m}} \mathcal{L}_{\lambda}(\mathbf{x}^{(t+1)},\mathbf{w}_{m}^{(t+1)},\mathbf{v}_{p}^{(t+1)},\mathbf{g}_{m},\mathbf{y}_{p}^{*},\mathbf{t}_{m}^{*} ).\label{Eq2:GP-ADMM:sub4}
			\end{align}
		\end{subequations}\par
		The problem\eqref{Eq2:GP-ADMM:sub2},\eqref{Eq2:GP-ADMM:sub3} and\eqref{Eq2:GP-ADMM:sub4} admit a closed-form solution and problem\eqref{Eq2:GP-ADMM:sub1} is solved using a GP step. Therefore, we next analyze the solution in problem\eqref{Eq2:GP-ADMM:sub1}. We first define the function
		\begin{equation}\label{key}
			\begin{split}
				&g(\mathbf{x};\mathbf{x}^{(t)},\mathbf{w}_{m}^{(t)},\mathbf{v}_{p}^{(t)},\mathbf{g}_{m}^{(t)},\mathbf{y}_{p}^{*},\mathbf{t}_{m}^{*})\\
				&=\mathcal{L}_{\lambda}(\mathbf{x}^{(t)},\mathbf{w}_{m}^{(t)},\mathbf{v}_{p}^{(t)},\mathbf{g}_{m}^{(t)},\mathbf{y}_{p}^{*},\mathbf{t}_{m}^{*})\\
				&+\Re\{\nabla_{\mathbf{x}}\mathcal{L}_{\lambda}(\mathbf{x}^{(t)},\mathbf{w}_{m}^{(t)},\mathbf{v}_{p}^{(t)},\mathbf{g}_{m}^{(t)},\mathbf{y}_{p}^{*},\mathbf{t}_{m}^{*})^{H}(\mathbf{x}-\mathbf{x}^{(t)})\}\\
				&+\dfrac{1}{2\alpha^{(t)}}\parallel \mathbf{x}-\mathbf{x}^{(t)}\parallel_{2}^{2},
			\end{split}
		\end{equation}
		where $ \alpha^{(t)} <\frac{1}{L^{(t)}}$, and $ L^{(t)} $ is the smallest Lipschitz constant of the function $ \mathcal{L}_{\lambda}(\mathbf{x},\mathbf{w}_{m},\mathbf{v}_{p},\mathbf{g}_{m},\mathbf{y}_{p},\mathbf{t}_{m}) $ at the $ t $-th iteration. According to the continuity of the function $ \mathcal{L}_{\lambda}(\mathbf{x},\mathbf{w}_{m},\mathbf{v}_{p},\mathbf{g}_{m},\mathbf{y}_{p},\mathbf{t}_{m}) $, we have the fact that
		\begin{equation}\label{key}
			\begin{split}
				&\mathcal{L}_{\lambda}(\mathbf{x},\mathbf{w}_{m}^{(t)},\mathbf{v}_{p}^{(t)},\mathbf{g}_{m}^{(t)},\mathbf{y}_{p}^{*},\mathbf{t}_{m}^{*})\\
				&\leq \mathcal{L}_{\lambda}(\mathbf{x}^{(t)},\mathbf{w}_{m}^{(t)},\mathbf{v}_{p}^{(t)},\mathbf{g}_{m}^{(t)},\mathbf{y}_{p}^{*},\mathbf{t}_{m}^{*})
				+L^{(t)}\parallel \mathbf{x}-\mathbf{x}^{(t)}\parallel_{2}^{2}.
			\end{split}
		\end{equation}
	    Then, we have
		\begin{equation}\label{key}
			\begin{split}
				&\mathcal{L}_{\lambda}(\mathbf{x},\mathbf{w}_{m}^{(t)},\mathbf{v}_{p}^{(t)},\mathbf{g}_{m}^{(t)},\mathbf{y}_{p}^{*},\mathbf{t}_{m}^{*})\\
				&\leq g(\mathbf{x};\mathbf{x}^{(t)},\mathbf{w}_{m}^{(t)},\mathbf{v}_{p}^{(t)},\mathbf{g}_{m}^{(t)},\mathbf{y}_{p}^{*},\mathbf{t}_{m}^{*}),\forall \mathbf{x}.
			\end{split}
		\end{equation}
		Furthermore, it can be seen that the function $ \mathcal{L}_{\lambda}(\mathbf{x},\mathbf{w}_{m},\mathbf{v}_{p},\mathbf{g}_{m},\mathbf{y}_{p}^{*},\mathbf{t}_{m}^{*} ) $ satisfies the following properties
		\begin{equation}\label{proof_L_g}
			\begin{split}
				&\mathcal{L}_{\lambda}(\mathbf{x}^{(t)},\mathbf{w}_{m}^{(t)},\mathbf{v}_{p}^{(t)},\mathbf{g}_{m}^{(t)},\mathbf{y}_{p}^{*},\mathbf{t}_{m}^{*})\\
				&= g(\mathbf{x}^{(t)};\mathbf{x}^{(t)},\mathbf{w}_{m}^{(t)},\mathbf{v}_{p}^{(t)},\mathbf{g}_{m}^{(t)},\mathbf{y}_{p}^{*},\mathbf{t}_{m}^{*}),
			\end{split}
		\end{equation}
		\begin{equation}\label{proof_L_g_GD}
			\begin{split}
				&\nabla_{\mathbf{x}}\mathcal{L}_{\lambda}(\mathbf{x}^{(t)},\mathbf{w}_{m}^{(t)},\mathbf{v}_{p}^{(t)},\mathbf{g}_{m}^{(t)},\mathbf{y}_{p}^{*},\mathbf{t}_{m}^{*})\\
				&=\nabla_{\mathbf{x}}g(\mathbf{x}^{(t)};\mathbf{x}^{(t)},\mathbf{w}_{m}^{(t)},\mathbf{v}_{p}^{(t)},\mathbf{g}_{m}^{(t)},\mathbf{y}_{p}^{*},\mathbf{t}_{m}^{*}).
			\end{split}
		\end{equation}
		Thus, the update of the variable $ \mathbf{x} $ can be expressed as
		\begin{equation}\label{proof_x}
			\mathbf{x}^{(t+1)}=\arg \min_{\mathbf{x}} g(\mathbf{x};\mathbf{x}^{(t)},\mathbf{w}_{m}^{(t)},\mathbf{v}_{p}^{(t)},\mathbf{g}_{m}^{(t)},\mathbf{y}_{p}^{*},\mathbf{t}_{m}^{*}).
		\end{equation}
		Now, considering the update of the problem\eqref{Eq2:GP-ADMM:sub2},\eqref{Eq2:GP-ADMM:sub3} and\eqref{Eq2:GP-ADMM:sub4}, we have the following inequality 
		\begin{subequations}\label{Eq3:GP-ADMM}
			\begin{align}
				&\mathcal{L}_{\lambda}(\mathbf{x}^{(t)},\mathbf{w}_{m}^{(t)},\mathbf{v}_{p}^{(t)},\mathbf{g}_{m}^{(t)},\mathbf{y}_{p}^{*},\mathbf{t}_{m}^{*})\label{Eq3:GP-ADMM:sub1}\\
				&= g(\mathbf{x}^{(t)};\mathbf{x}^{(t)},\mathbf{w}_{m}^{(t)},\mathbf{v}_{p}^{(t)},\mathbf{g}_{m}^{(t)},\mathbf{y}_{p}^{*},\mathbf{t}_{m}^{*})\label{Eq3:GP-ADMM:sub2}\\
				&\geq g(\mathbf{x}^{(t)};\mathbf{x}^{(t+1)},\mathbf{w}_{m}^{(t)},\mathbf{v}_{p}^{(t)},\mathbf{g}_{m}^{(t)},\mathbf{y}_{p}^{*},\mathbf{t}_{m}^{*})\label{Eq3:GP-ADMM:sub3}\\
				&\geq 	\mathcal{L}_{\lambda}(\mathbf{x}^{(t+1)},\mathbf{w}_{m}^{(t)},\mathbf{v}_{p}^{(t)},\mathbf{g}_{m}^{(t)},\mathbf{y}_{p}^{*},\mathbf{t}_{m}^{*})\label{Eq3:GP-ADMM:sub4}\\
				&\geq 	\mathcal{L}_{\lambda}(\mathbf{x}^{(t+1)},\mathbf{w}_{m}^{(t+1)},\mathbf{v}_{p}^{(t)},\mathbf{g}_{m}^{(t)},\mathbf{y}_{p}^{*},\mathbf{t}_{m}^{*})\label{Eq3:GP-ADMM:sub5}\\
				&\geq 	\mathcal{L}_{\lambda}(\mathbf{x}^{(t+1)},\mathbf{w}_{m}^{(t+1)},\mathbf{v}_{p}^{(t+1)},\mathbf{g}_{m}^{(t)},\mathbf{y}_{p}^{*},\mathbf{t}_{m}^{*})\label{Eq3:GP-ADMM:sub6}\\
				&\geq 	\mathcal{L}_{\lambda}(\mathbf{x}^{(t+1)},\mathbf{w}_{m}^{(t+1)},\mathbf{v}_{p}^{(t+1)},\mathbf{g}_{m}^{(t+1)},\mathbf{y}_{p}^{*},\mathbf{t}_{m}^{*}),\label{Eq3:GP-ADMM:sub7}
			\end{align}
		\end{subequations}
		where\eqref{Eq3:GP-ADMM:sub2} follows from\eqref{proof_L_g},\eqref{Eq3:GP-ADMM:sub3} holds because of\eqref{proof_x} and\eqref{Eq2:GP-ADMM:sub1},\eqref{Eq3:GP-ADMM:sub5},\eqref{Eq3:GP-ADMM:sub6} and\eqref{Eq3:GP-ADMM:sub7} are optimal solutions obtained by solving the subproblems\eqref{Eq2:GP-ADMM:sub2},\eqref{Eq2:GP-ADMM:sub3} and\eqref{Eq2:GP-ADMM:sub4}, respectively.\par
		Further, we have
		\begin{equation}\label{proof_objective function_lim}
			\begin{split}
				&\lim_{t\to\infty}\mathcal{L}_{\lambda}(\mathbf{x}^{(t+1)},\mathbf{w}_{m}^{(t+1)},\mathbf{v}_{p}^{(t+1)},\mathbf{g}_{m}^{(t+1)},\mathbf{y}_{p}^{*},\mathbf{t}_{m}^{*})\\
				=&\lim_{t\to\infty} f(\mathbf{x}^{(t+1)},\{\mathbf{w}_{m}^{(t+1)}\})\\
				&+\lim_{t\to\infty}\dfrac{\lambda}{2}\sum_{p=1}^{M_{\text{jam}}} \parallel\mathbf{v}_{p}^{(t+1)}-\mathbf{G}^{H}(\theta_{\text{jam},p})\mathbf{x}^{(t+1)}+\mathbf{s}_{\text{jam},p}\parallel_{2}^{2}\\
				&+\lim_{t\to\infty}\lambda\Re\{(\mathbf{y}_{p}^{*})^{H}(\mathbf{v}_{p}^{(t+1)}-\mathbf{G}^{H}(\theta_{\text{jam},p})\mathbf{x}^{(t+1)}+\mathbf{s}_{\text{jam},p})\}\\
				&+\lim_{t\to\infty}\dfrac{\lambda}{2}\sum_{m=1}^{M} \parallel\mathbf{g}_{m}^{(t+1)}-\mathbf{T}_{m}^{H}\mathbf{x}^{(t+1)}\parallel_{2}^{2}\\
				&+\lim_{t\to\infty}\lambda\Re\{(\mathbf{t}_{m}^{*})^{H}(\mathbf{g}_{m}^{(t+1)}-\mathbf{T}_{m}^{H}\mathbf{x}^{(t+1)})\}\\
				=&\lim_{t\to\infty} f(\mathbf{x}^{(t+1)},\{\mathbf{w}_{m}^{(t+1)}\}).
			\end{split}
		\end{equation}
		Since $ f(\mathbf{x}^{(t+1)},\{\mathbf{w}_{m}^{(t+1)}\})>0 $, the objective function $ f(\mathbf{x}^{(t+1)},\{\mathbf{w}_{m}^{(t+1)}\}) $ has lower bound. According to\eqref{Eq3:GP-ADMM} and\eqref{proof_objective function_lim}, it can be seen that the objective function converges, which proves that the proposed GP-ADMM algorithm converges when it satisfies $\alpha^{(t)} <\frac{1}{L^{(t)}}$. $ \hfill\blacksquare $\par

		\section{proof of Lemma \ref{Lemma1}}\label{appendixB}
		$ Proof $: Assume that $ \{t_{j}\} $ signifies the index set of a convergent subsequence, and that $\{\mathbf{x},\mathbf{w}_{m},\mathbf{v}_{p},\mathbf{g}_{m},\mathbf{y}_{p},\mathbf{t}_{m} \}$ converges to $ \{\mathbf{x}^{*},\mathbf{w}_{m}^{*},\mathbf{v}_{p}^{*},\mathbf{g}_{m}^{*},\mathbf{y}_{p}^{*},\mathbf{t}_{m}^{*} \} $. Then,
		we have
		\begin{subequations}\label{Eq4:GP-ADMM}
			\begin{align}
				&g(\mathbf{x};\mathbf{x}^{(t_{j})},\mathbf{w}_{m}^{(t_{j})},\mathbf{v}_{p}^{(t_{j})},\mathbf{g}_{m}^{(t_{j})},\mathbf{y}_{p}^{*},\mathbf{t}_{m}^{*})\label{Eq4:GP-ADMM:sub1}\\
				&\geq 	g(\mathbf{x}^{(t_{j}+1)};\mathbf{x}^{(t_{j})},\mathbf{w}_{m}^{(t_{j})},\mathbf{v}_{p}^{(t_{j})},\mathbf{g}_{m}^{(t_{j})},\mathbf{y}_{p}^{*},\mathbf{t}_{m}^{*})\label{Eq4:GP-ADMM:sub2}\\
				&\geq \mathcal{L}_{\lambda}(\mathbf{x}^{(t_{j}+1)},\mathbf{w}_{m}^{(t_{j})},\mathbf{v}_{p}^{(t_{j})},\mathbf{g}_{m}^{(t_{j})},\mathbf{y}_{p}^{*},\mathbf{t}_{m}^{*})\label{Eq4:GP-ADMM:sub3}\\
				&\geq \mathcal{L}_{\lambda}(\mathbf{x}^{(t_{j}+1)},\mathbf{w}_{m}^{(t_{j}+1)},\mathbf{v}_{p}^{(t_{j})},\mathbf{g}_{m}^{(t_{j})},\mathbf{y}_{p}^{*},\mathbf{t}_{m}^{*})\label{Eq4:GP-ADMM:sub4}\\
				&\geq \mathcal{L}_{\lambda}(\mathbf{x}^{(t_{j}+1)},\mathbf{w}_{m}^{(t_{j}+1)},\mathbf{v}_{p}^{(t_{j}+1)},\mathbf{g}_{m}^{(t_{j})},\mathbf{y}_{p}^{*},\mathbf{t}_{m}^{*})\label{Eq4:GP-ADMM:sub5}\\
				&\geq \mathcal{L}_{\lambda}(\mathbf{x}^{(t_{j}+1)},\mathbf{w}_{m}^{(t_{j}+1)},\mathbf{v}_{p}^{(t_{j}+1)},\mathbf{g}_{m}^{(t_{j}+1)},\mathbf{y}_{p}^{*},\mathbf{t}_{m}^{*})\label{Eq4:GP-ADMM:sub6}\\
				&\geq \mathcal{L}_{\lambda}(\mathbf{x}^{(t_{j+1})},\mathbf{w}_{m}^{(t_{j+1})},\mathbf{v}_{p}^{(t_{j+1})},\mathbf{g}_{m}^{(t_{j+1})},\mathbf{y}_{p}^{*},\mathbf{t}_{m}^{*})\label{Eq4:GP-ADMM:sub7}\\
				&=g(\mathbf{x}^{(t_{j}+1)};\mathbf{x}^{(t_{j+1})},\mathbf{w}_{m}^{(t_{j+1})},\mathbf{v}_{p}^{(t_{j+1})},\mathbf{g}_{m}^{(t_{j+1})},\mathbf{y}_{p}^{*},\mathbf{t}_{m}^{*}),\label{Eq4:GP-ADMM:sub8}
			\end{align}
		\end{subequations}
		where\eqref{Eq4:GP-ADMM:sub6} holds because of $ t_{j} $ is an index of a subsequence, and $ t_{j+1}\geq t_{j}+1 $. Next, utilizing the continuity of the function $ g $ and taking the limit $ j\to \infty $, we obtain
		\begin{equation}\label{proof_KKT_g}
			g(\mathbf{x};\mathbf{x}^{*},\mathbf{w}_{m}^{*},\mathbf{v}_{p}^{*},\mathbf{g}_{m}^{*},\mathbf{y}_{p}^{*},\mathbf{t}_{m}^{*})\geq g(\mathbf{x}^{*};\mathbf{x}^{*},\mathbf{w}_{m}^{*},\mathbf{v}_{p}^{*},\mathbf{g}_{m}^{*},\mathbf{y}_{p}^{*},\mathbf{t}_{m}^{*}).
		\end{equation}
		The inequality in\eqref{proof_KKT_g} means that $ \mathbf{x}^{*} $ is a blockwise minimizer
		of  the function $ g(\mathbf{x};\mathbf{x}^{*},\mathbf{w}_{m}^{*},\mathbf{v}_{p}^{*},\mathbf{g}_{m}^{*},\mathbf{y}_{p}^{*},\mathbf{t}_{m}^{*}) $, and it satisfies the following partial KKT condition on $ \mathbf{x} $
		\begin{equation}\label{proof_KKT_x}
			\nabla_{\mathbf{x}}g(\mathbf{x}^{*};\mathbf{x}^{*},\mathbf{w}_{m}^{*},\mathbf{v}_{p}^{*},\mathbf{g}_{m}^{*},\mathbf{y}_{p}^{*},\mathbf{t}_{m}^{*})+2\varrho^{*}\mathbf{x}^{*}=\mathbf{0}_{N_{T}N\times 1},
		\end{equation}
		where $\varrho^{*}\in\mathbb{R}^{+} $ denotes the optimal  dual solution associated with the constraint set $ \bm{\Omega} $. According to\eqref{proof_L_g_GD} we have that
		\begin{equation}\label{key}
			\nabla_{\mathbf{x}}\mathcal{L}_{\lambda}(\mathbf{x}^{*},\mathbf{w}_{m}^{*},\mathbf{v}_{p}^{*},\mathbf{g}_{m}^{*},\mathbf{y}_{p}^{*},\mathbf{t}_{m}^{*})+2\varrho^{*}\mathbf{x}^{*}=\mathbf{0}_{N_{T}N\times 1}.
		\end{equation}
		Similarly, according to the update rule of\eqref{Eq2:GP-ADMM:sub2}, we have
		\begin{equation}\label{key}
			\begin{split}
				&\mathcal{L}_{\lambda}(\mathbf{x}^{(t_{j}+1)},\mathbf{w}_{m},\mathbf{v}_{p}^{(t_{j})},\mathbf{g}_{m}^{(t_{j})},\mathbf{y}_{p}^{*},\mathbf{t}_{m}^{*})\\
				&\geq
				\mathcal{L}_{\lambda}(\mathbf{x}^{(t_{j}+1)},\mathbf{w}_{m}^{(t_{j}+1)},\mathbf{v}_{p}^{(t_{j})},\mathbf{g}_{m}^{(t_{j})},\mathbf{y}_{p}^{*},\mathbf{t}_{m}^{*}),
			\end{split}
		\end{equation}
		and thus
		\begin{equation}\label{key}
		   \mathcal{L}_{\lambda}(\mathbf{x}^{*},\mathbf{w}_{m},\mathbf{v}_{p}^{*},\mathbf{g}_{m}^{*},\mathbf{y}_{p}^{*},\mathbf{t}_{m}^{*})\geq
		   \mathcal{L}_{\lambda}(\mathbf{x}^{*},\mathbf{w}_{m}^{*},\mathbf{v}_{p}^{*},\mathbf{g}_{m}^{*},\mathbf{y}_{p}^{*},\mathbf{t}_{m}^{*}).
		\end{equation}
		Then, $ \mathbf{w}_{m}^{*} $ is a blockwise minimizer of the function $ \mathcal{L}_{\lambda}(\mathbf{x}^{*},\mathbf{w}_{m},\mathbf{v}_{p}^{*},\mathbf{g}_{m}^{*},\mathbf{y}_{p}^{*},\mathbf{t}_{m}^{*}) $, and satisfies the following partial KKT condition on $ \mathbf{w}_{m} $ 
		\begin{equation}\label{proof_KKT_w}
			\nabla_{\mathbf{w}_{m}}\mathcal{L}_{\lambda}(\mathbf{x}^{*},\mathbf{w}_{m}^{*},\mathbf{v}_{p}^{*},\mathbf{g}_{m}^{*},\mathbf{y}_{p}^{*},\mathbf{t}_{m}^{*})=\mathbf{0}_{N_{R}N\times 1}.
		\end{equation}
		
		For the same reason, we can obtain partial KKT conditions on $ \mathbf{v}_{p} $ and $ \mathbf{g}_{m} $ by the update rule of\eqref{Eq2:GP-ADMM:sub3} and\eqref{Eq2:GP-ADMM:sub4}
		\begin{equation}\label{proof_KKT_v_p}
			\nabla_{\mathbf{v}_{p}}\mathcal{L}_{\lambda}(\mathbf{x}^{*},\mathbf{w}_{m}^{*},\mathbf{v}_{p}^{*},\mathbf{g}_{m}^{*},\mathbf{y}_{p}^{*},\mathbf{t}_{m}^{*})=\mathbf{0}_{N\times 1}.
		\end{equation}
	    \begin{equation}\label{proof_KKT_g_m}
	    	\nabla_{\mathbf{g}_{m}}\mathcal{L}_{\lambda}(\mathbf{x}^{*},\mathbf{w}_{m}^{*},\mathbf{v}_{p}^{*},\mathbf{g}_{m}^{*},\mathbf{y}_{p}^{*},\mathbf{t}_{m}^{*})=\mathbf{0}_{N\times 1}.
	    \end{equation}
		Finally, combining the KKT conditions\eqref{proof_KKT_x},\eqref{proof_KKT_w},\eqref{proof_KKT_v_p} and\eqref{proof_KKT_g_m}, we show that the sequence of solutions generated by the GP-ADMM algorithm converges to the KKT point of the problem\eqref{Eq1:GP-ADMM}. The proof is completed. $ \hfill\blacksquare $\par


\end{document}